\documentclass[review,3p]{elsarticle}

\usepackage[utf8x]{inputenc}
\usepackage{graphicx}
\usepackage{rotating}
\usepackage{subcaption}
\usepackage{amssymb}
\usepackage{amsmath}
\usepackage{lineno}
\usepackage{amsmath}
\usepackage{booktabs}
\usepackage{gensymb}
\usepackage{listings}
\usepackage{xcolor}
\usepackage{multirow}
\usepackage{lscape}
\usepackage{multicol}
\usepackage{soul}
\usepackage{cuted}
\usepackage{flushend}
\usepackage{url}
\usepackage[colorlinks=True]{hyperref}

\biboptions{authoryear}

\usepackage{array}
\newcolumntype{L}[1]{>{\raggedright\let\newline\\\arraybackslash\hspace{0pt}}m{#1}}
\newcolumntype{C}[1]{>{\centering\let\newline\\\arraybackslash\hspace{0pt}}m{#1}}
\newcolumntype{R}[1]{>{\raggedleft\let\newline\\\arraybackslash\hspace{0pt}}m{#1}}

\def\equationautorefname~#1\null{(#1)\null}

\journal{Renewable Energy}

\begin{document}

\begin{frontmatter}

\title{A parametric model for wind turbine power curves incorporating environmental conditions}

\author[MinesParistech]{Yves-Marie Saint-Drenan\corref{correspondingauthor1}}
\cortext[correspondingauthor1]{Corresponding author}
\ead{yves-marie.saint-drenan@mines-paristech.fr} 
\author[MinesParistech]{Romain Besseau} 
\author[ImpCollege]{Malte Jansen}
\author[ImpCollege]{Iain Staffell} 
\author[UEA,WEMC]{Alberto Troccoli} 
\author[EDF,WEMC]{Laurent Dubus} 
\author[BOKU]{Johannes Schmidt}
\author[BOKU]{Katharina Gruber}
\author[CENSE]{Sofia G. Simões}
\author[UniKS]{Siegfried Heier}

\address[MinesParistech]{MINES ParisTech, PSL Research University, O.I.E. Centre Observation, Impacts, Energy, 06904 Sophia Antipolis, France}
\address[ImpCollege]{Centre for Environmental Policy, Imperial College London, London SW7 1NE, UK}
\address[EDF]{EDF RD/MFEE, Applied Meteorology and Atmospheric Environment, CHATOU CEDEX, France}
\address[UEA]{School of Environmental Sciences, University of East Anglia, Norwich, NR4 7TJ, UK}
\address[WEMC]{World Energy and Meteorology Council (WEMC), Norwich, NR4 7TJ, UK}
\address[BOKU]{Institute for Sustainable Economic Development, University of Natural Resources and Life Sciences, 1190 Vienna, Austria } 
\address[CENSE]{CENSE – Center for Environmental and Sustainability Research, NOVA School for Science and Technology, NOVA University Lisbon, 2829-516 Caparica, Portugal}
\address[UniKS]{University of Kassel, Kassel, Germany}

\begin{abstract} 
A wind turbine's power curve relates its power production to the wind speed it experiences.  The typical shape of a power curve is well known and has been studied extensively; however, the power curves of individual turbine models can vary widely from one another.  This is due to both the technical features of the turbine (power density, cut-in and cut-out speeds, limits on rotational speed and aerodynamic efficiency), and environmental factors (turbulence intensity, air density, wind shear and wind veer).  Data on individual power curves are often proprietary and only available through commercial databases.  

We therefore develop an open-source model which can generate the power curve of any turbine, adapted to the specific conditions of any site.  This can employ one of six parametric models advanced in the literature, and accounts for the eleven variables mentioned above.  The model is described, the impact of each technical and environmental feature is examined, and it is then validated against the manufacturer power curves of 91 turbine models.  Versions of the model are made available in MATLAB, R and Python code for the community.

\end{abstract}


\end{frontmatter}

\section{Introduction}\label{sec:Introduction} 
The power curve of a wind turbine relates the speed of the wind flow intercepted by the wind turbine rotor to its electrical output. 
A power curve is needed at different stages of the lifetime of a wind farm. 
Prior to its market introduction, the power curve of a newly designed turbine must be assessed to validate its performance. 
Project developers use power curves together with wind information to evaluate the economic viability of developing a wind farm. 
When operating, the aerodynamic efficiency of a turbine may evolve over time due to wear of turbine components, dirt accumulation on the wind turbine blades, and many other effects \citep{Brown_2012}. 
Evaluation of the power curve during the lifetime of a wind farm is therefore useful to monitor the state of health of the turbines \citep{Brown_2012} and degradation due to ageing \citep{STAFFELL2014775,SHIN20171180,DAI2018199}. 
Power curves are also used to estimate the aggregated power production of wind farms, and their integration into national power systems and electricity markets \citep{GonzalezAparicio2016,Staffell2016}.

There has been extensive research on methods for assessing power curves over the last decades \citep{Elliott1990,Sumner2006,Kaiser2007,Gottschall2008,Rareshide2009,Wagner2009,Albers2010,Wharton2012}. 
Indeed, the quality of power curves is a critical issue since the risk involved in the building and operation of wind farms depends directly on the accuracy of this information. 
Ideally, power curves should be measured in a wind tunnel under controlled conditions. Due to the large dimensions of modern wind turbines, power curves can only be evaluated in real outdoor conditions, making robust assessment difficult due to the spatial and temporal variations of the wind speed. 
In addition, measurement procedures recommended in the IEC standard 61400-12-1 \citep{IEC2005} are continuously improved \citep{PCWG}. 
All these activities jointly contribute to the increased accuracy of assessed power curve and in a reduction of the acquisition time needed to evaluate them.

Power curves are assessed and made available by turbine manufacturers after the correction of different issues, such as turbulence intensity, wind shear, wind veer, up-flow angle; following the procedures defined by the IEC standard 61400-12-1 \citep{IEC2005}. 
These power curves can be found in the product sheets of wind turbines or in databases which collate numerous power curves, such as  \citet{thewindpower} or WindPRO \citep{WindPRO}, which are used in this study. While convenient, these databases are not freely available.
Unfortunately, power curves of many wind turbines remain difficult to find and, when available, information such as the reference turbulence intensity or the air density is frequently missing. 
This lack of information leads to a non-negligible uncertainty in the power calculation of a given turbine at best, and to the impossibility of performing such calculation at worst. 
It is particularly an issue in prospective analyses of the energy mix where power curves are required for each individual turbine installed across a wide area \citep{GonzalezAparicio2016,Staffell2016}. 
This paper addresses these issues mentioned above: the availability of power curves and the consideration of environmental parameters, by proposing a parameterised power curve model where the impact of turbulence intensity, wind shear, wind veer and air density are explicitly considered.

There are different possibilities for estimating the power production from wind speed data when the power curve is unknown. 
One approach consists of using a statistical model whose parameters are trained on joint measurements of power output and meteorological inputs for a historical period. 
An impressive number of analytical and statistical tools have been identified, including polynomial models, linearised segmented models, neural networks and fuzzy methods \citet{LYDIA2014,Sohoni2016}. 

If statistical models look similar to a power curve at first glance and can be exchanged in some applications, they differ on some important points. 
Firstly, statistical models capture the relationship between wind speed and net (rather than gross) power production, including potential wake effects, the impact of the local orography, wind turbine availability or even systematic errors in the wind speed data.  These factors should be disentangled as the gross turbine production is of interest. 
Secondly, supervised statistical models require training data and therefore they cannot be used to model a planned wind farm or to simulate a fleet of wind farms where available measurements do not yet exist. 
In the latter case, the use of power curve is unavoidable and the lack of information is often addressed by choosing equivalent power curves based on the similarity between the desired turbine and those for which a power curve is available \citep{BECKER2017252}. 
However, even this approach lacks a widely recognised and validated rationale.

As described in numerous studies on the dynamics of wind turbines \citep{Heier2014,Tian2017}, the behaviour of the power production of a turbine can be estimated as a function of the wind speed using general characteristics of the turbine and a power coefficient model. 
To the best of our knowledge, the use of such models for generating power curves has not been systematically studied and validated so far. 
The physically-based approach suggested here uses the rated power and the rotor dimension as the main input parameters, and allows other operating characteristics, which are also standard information, to also be specified (such as cut-in and cut-out wind speed, or minimal or maximal rotational speed). 
This work relies on existing analytic power coefficient functions describing the aerodynamic efficiency of the blade published in the literature such as e.g. \citep{Heier2014,Dai2016} but other input data stemming from blade measurements or numerical calculation can be used instead. 
Finally, the model proposed here offers the possibility to explicitly account for environmental factors such as turbulence intensity, air density, wind shear and veer.

This paper is structured in six main parts. 
A comprehensive description is given in \autoref{sec:Methodology} of the different steps necessary to evaluate a power curve from general characteristics of a wind farm, such as the rotor area or the nominal power.  
This section also includes a discussion on the consideration and influence of external environmental parameters. 
Owing to the numerous input parameters of the model, a sensitivity analysis and statistical analysis of these parameters are described in \autoref{sec:sensitivity} and \autoref{sec:StatisticalAnalysis}, respectively. 
The results of our validation are summarised in \autoref{sec:Validation}, where the model output has been compared to power curve from \citet{thewindpower} database. 
Some insights on the limitations and possible improvements to the proposed model are discussed in \autoref{sec:Conclusion}, as well as on its possible domains of application. 
Implementations of the proposed model in Python, R and MATLAB are also provided as supplementary material to this paper.

\section{Methodology}\label{sec:Methodology}
\subsection{Operating regions of a wind turbine}\label{sec:OperatingRegions}

\begin{figure}[h!]
 \centering
    \includegraphics[width=0.8\textwidth]{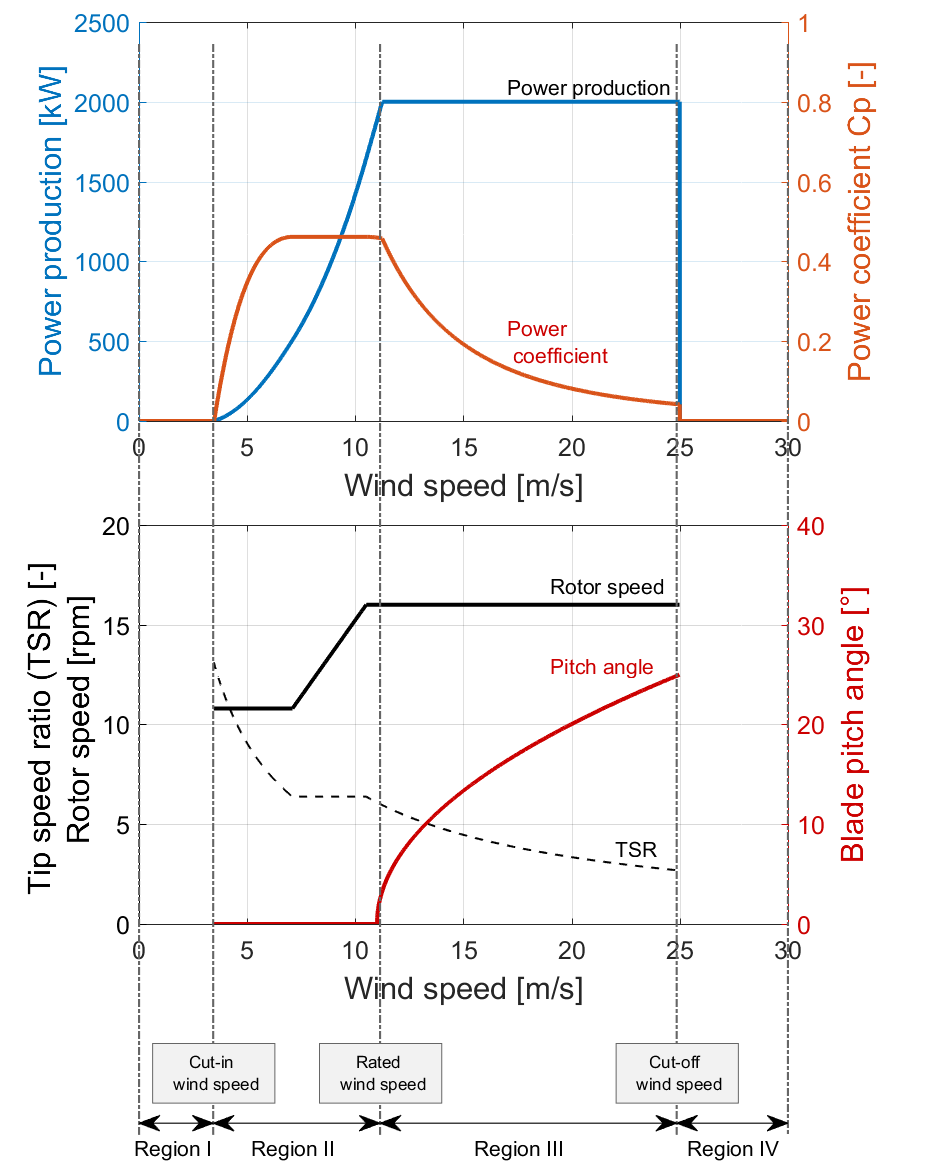}
    \caption{Operating regions of a wind turbine and evolution of pitch angle, rotation speed and tip-speed ratio (TSR) with wind speed}
    \label{fig:WTOperatingRegions}
\end{figure}

Power curves are traditionally divided into four operation regions, as shown in \autoref{fig:WTOperatingRegions} and detailed below. 
At very low wind speeds, the torque exerted by the wind on the blades is insufficient to bring the turbine to rotate. 
The wind speed at which the turbine starts to generate electricity is called cut-in wind speed and is typically between 3 and 4 m/s. 
Region I corresponds to wind speeds below this cut-in wind speed.  Power can be consumed in this region from turbine electronics, communications and heating / de-icing of blades, although these ancillary loads are not included in power curves.

Above the cut-in wind speed, there is sufficient torque for rotation, and power production increases with the cube of wind speed before reaching a threshold corresponding to the rated power of the turbine (or nominal power) that is designed to not exceed. 
The lowest wind speed at which the nominal power is reached is called the rated (or nominal) wind speed and is typically between 12 and 17 m/s.
Region II is delimited by the cut-in and the rated wind speed, and corresponds to an interval where the wind turbine operates at maximal efficiency. 
There are, however, some exceptions to the optimal operation of the wind turbine in this region. 
Firstly, while an optimal operation requires the rotational speed to be proportional to the wind speed, the speed of rotation is bounded by lower and upper limits. 
Secondly, at high wind speeds, the turbine can sometimes be deliberately operated at lower power to reduce rotor torque and noise levels \citep{Luo2017}.

For wind speeds above the rated wind speed, the wind turbine is designed to keep output power at the rated power, which cannot be exceeded. 
This can be achieved by means of a stall regulation or pitch control. 
The latter solution consists in adjusting the pitch angle of the blades to keep the power at the constant level and is overwhelmingly used in modern large turbines. 
Region III corresponds to wind speed values where the turbine operates at its rated power, and is bounded by the nominal wind speed and the cut-off wind speed, which is introduced below.

The forces acting on the turbine structure increase with wind speed, and at some point the structural condition of the turbine can be endangered. 
To prevent damage, a braking system is employed to bring the rotor to a standstill \citep{Wood1996}. 
The cut-off wind speed corresponds to the maximum wind speed a wind turbine can safely support while generating power and is usually about 25 m/s. 
Region IV includes all wind speeds larger than the cut-off wind speed.  Some manufacturers have introduced storm control in larger-bladed turbine models, where the power is gradually reduced (e.g. from 21 m/s up to 25 m/s) to prevent such drastic loss of power at the cut-out speed. 

\autoref{fig:WTOperatingRegions} shows the different operating regions described above as well as the evolution of the main operating parameters of a wind turbine: pitch angle, rotor speed and tip-speed ratio (TSR). This visual representation is based on previous works \citep{Kvittem2012,Campagnolo2016,Avossa2017,Dai2016}.

\subsection{Parametric wind turbine power curve}\label{sec:ParametricPowerCurve}
   
The wind power calculation for regions I, III and IV is trivial with the information typically available on a wind turbine\footnote{The cut-in, cut-off and rated wind speed as well as the rated power are typically given in wind turbine product sheets. If not available, missing parameters can still be estimated, as explained later.}. 
However, the description of the power curve in region II is complex and methodologies to improve it are still being researched.
Between the cut-in and the rated wind speeds, the wind power production $P_{WT}$ can be calculated by Eq. \autoref{eq:1} as a function of the wind speed $V_{WS}$, air density $\rho$, rotor area $A_{rotor}$ and power coefficient $C_{p}(\lambda,\beta)$, with $\lambda$ being the tip-speed ratio and $\beta$ the blade angle \citep{Heier2014}: 

\begin{equation}
\label{eq:1}
P_{WT}=\frac{1}{2} \rho A_{rotor} V_{WS}^{3} C_{p} (\lambda , \beta )
\end{equation}
\bigbreak

Rotor area is straightforward to obtain and data on air density are readily available, although it should be noted that density at a specific site varies over time for a specific site (for example between 1.1 kg/m\textsuperscript{3} and 1.3 kg/m\textsuperscript{3} between summer and winter in Germany \citep{PFENNINGER20161251}.
That said, the parameter with the largest uncertainty in Eq.~\autoref{eq:1} is the power coefficient $C_{p}(\lambda,\beta)$t which ultimately depends on the wind speed. 

\bigbreak
\textbf{Parametric model of the power coefficient $C_{p} (\lambda , \beta )$}

The power coefficient $C_{p}(\lambda,\beta)$ expresses the recoverable fraction of the power in the wind flow. 
This quantity is generally assumed to be a function of both tip-speed ratio $\lambda$ and blade pitch angle $\beta$ \citep{Heier2014}. 
The power coefficient can either be evaluated experimentally or calculated numerically using blade element momentum (BEM), computational fluid dynamics (CFD) or generalised dynamic wake (GDW) models \citep{Slootweg2001,Dai2012,Dai2016}. 
A less accurate but convenient alternative consists in using numerical approximations.
A few empirical relations can be found in the literature (see e.g. \citep{Heier2014}) with the general form: 

\begin{equation}
\label{eq:2}
       \begin{cases}
          C_{p}(\lambda,\beta)=c_{1}(c_{2}/\lambda_{i}-c_{3}\beta-c_{4}\lambda_{i}\beta-c_{5}\beta^{x}-c_{6})e^{-c_{7}/\lambda_{i}}+c_{8}\lambda\\
          \lambda_{i}^{-1}=(\lambda+c_{9} \beta)^{-1}-c_{10}(\beta^{3}+1)^{-1}
       \end{cases}
\end{equation}
\bigbreak
The above equation reflects the general relationship of the power coefficient $C_{p}$ with $\lambda$ and $\beta$. 
Different values can be found for the model coefficients $c_{i}$. 
In our study, six different parameter sets from different authors have been considered \citep{Slootweg2003,Thongam2009,DeKooning2013,Ochieng2014,Dai2016}. 
These different parameterisations are listed and illustrated in \ref{appendix:Annex1}. 
We limit the extent of the analysis to six parameterisations but our approach can be easily extended to any other parametric models or numerical data. 

\bigbreak
\textbf{Determination of the blade pitch angle, $\beta$, as a function of wind speed}

If we assume that the wind turbine is designed to achieve its maximum efficiency in region II, the blade pitch angle can be set to zero between the cut-in and the nominal wind speed. 
Indeed, it is usually assumed that the blade pitch is only used to limit the power production to the nominal power in region III \citep{Dai2016,Avossa2017,Luo2017} and our modelling assumption seems therefore reasonable. 
That said, pitch angle can be used in regulation strategies that aim to limit noise emissions or mechanical effects on the turbine structure within region II as well \citep{Luo2017}. Such strategies are not considered in the present work and their integration in our modelling approach may be the subject of future developments. 
\bigbreak
\textbf{Determination of the tip-speed ratio $\lambda$ as a function of the wind speed}

The tip-speed of the blade is equal to the product of the rotational speed of the rotor $\omega$ and the rotor radius, $D_{rotor}/2$. 
We can therefore express the tip-speed ratio as a function of the rotor rotational speed and radius as well as of the wind speed $V_{WS}$ as follows: 

\begin{equation}
\label{eq:3}
       \lambda=\frac{\omega \cdot (D_{rotor}/2)}{V_{WS}}
\end{equation}

As explained above, the aerodynamic efficiency of the wind turbine depends on the tip-speed ratio $\lambda$. 
The maximum power yield in region II is therefore obtained for $\lambda_{opt}$ namely the value that maximises $C_{p}$ for a given wind speed: 

\begin{equation}
\label{eq:4}
       \lambda_{opt}=\underset{\lambda,\beta=0}{\operatorname{arg \, max}} \: C_{p}(\lambda,\beta)
\end{equation}

Considering Eq. \autoref{eq:3}, if the wind turbine is operating at constant tip-speed ratio, the rotational speed of the rotor $\omega$ should vary proportionally to the wind speed $V_{WS}$. 
This is only possible in the operating range of the turbine, which is bounded by $\omega_{min}$ and $\omega_{max}$. 
This constraint should be taken into account in the estimation of $\lambda$ according to Eq.~\autoref{eq:3}. 
Based on previous works (e.g. \citep{Kvittem2012,Avossa2017}), we use a simple approach, which consists in estimating the value of $\lambda$ using the rotational speed $\omega$ as follows:

\begin{equation}
\label{eq:5}
       \omega=min \left(\omega_{max},max \left( \omega_{min},\frac{\lambda_{opt}}{D_{rotor}/2} \cdot V_{WS} \right ) \right)
\end{equation}

As illustrated in \autoref{fig:WTOperatingRegions}, the rotational speed $\omega$ given by Eq. \autoref{eq:5} corresponds to  $\lambda_{opt}$ but is bounded between $\omega_{min}$ and $\omega_{max}$.
It can be observed that the maximum value of $C_{p}$ with constrained rotational speed $\omega$ is obtained with Eq. \autoref{eq:5} due to the monotonic behaviour of the function $C_{p}(\lambda)$ for $\lambda<\lambda_{opt}$ and $\lambda>\lambda_{opt}$ respectively.

\subsection{Considering the effect of external parameters on the power curve}\label{sec:ExternalParam}

As summarised later in \autoref{sec:SummaryModel}, the relationships and modelling assumptions described in \autoref{sec:ParametricPowerCurve} are sufficient for estimating the power curve of a wind turbine. 
However, this power curve corresponds to ideal operating conditions and external factors should be taken into consideration to better simulate the behaviour of a wind turbine in real conditions. 
These factors are the turbulence intensity, air density, wind shear and wind veer, inflow angle and wake effects.

Wake effects are strongly dependent on the specific layout of a wind farm, particularly the number of turbines and their spacing. 
Wake losses amount to approximately 11 to 13 \% for turbines spaced 7 to 9 turbine diameters apart \citep{GONZALEZLONGATT2012329,BOSCH2018766}. 
As the losses are time-varying, due to wind speed and its prevailing direction, they can not be considered further in the present work. 
The inflow angle results from the effect of the orography on the wind but, since it depends on the site and less on the wind turbine itself, it is not considered here. 
The effects of the remaining parameters on the power curve are evaluated next. 

\bigbreak
\textbf{The effect of turbulence intensity on the power curve}

The power curve derived in the previous section corresponds to the ideal case of a laminar and stationary wind conditions, which rarely occurs in practice. 
Since the relationship between wind power and wind speed is non-linear, the effect of high frequency variations in the wind speed on the power must be taken into consideration \citep{Nrgaard2004AMP}. 
This is usually realised by considering the turbulence intensity (TI) defined as:

\begin{equation}
\label{eq:6}
       TI=\frac{\sigma(u)}{\mu(u)}
\end{equation}

In the above equation, $\mu(u)$ represents the mean wind speed and $\sigma(u)$ the standard deviation of the wind speed measured at a frequency of 1Hz or higher in a time period of 10 minutes \citep{IEC2005}.
Typical values for the average turbulence intensity range from 5 to 15~\%. 
When no time series of the turbulence intensity is available, it is usual to assume a constant value of the turbulence intensity for a particular site.

\begin{figure}[h!]
 \centering
    \includegraphics[width=0.8\textwidth]{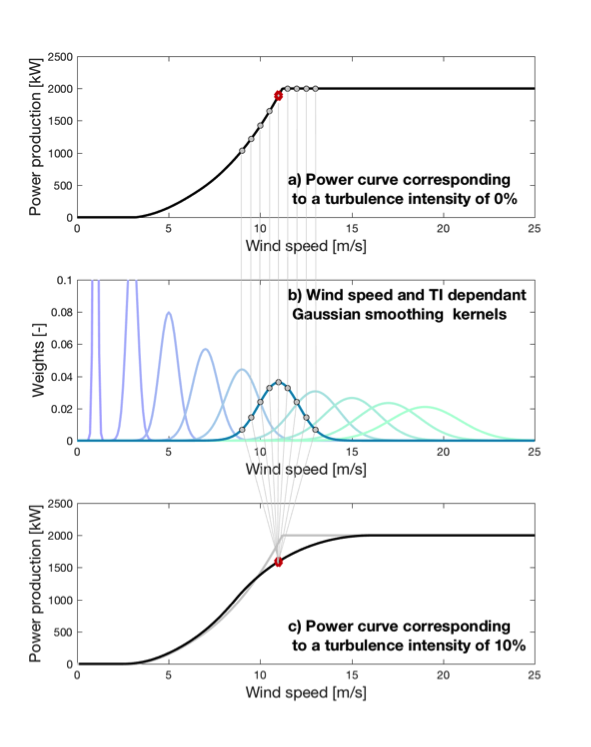}
    \caption{Illustration of the method used to calculate the effect of turbulence intensity on the power curve: the upper, middle and lower plots represent respectively the original power curve, the different Kernels and the final power curve. An example of calculation for a wind speed of 11 m/s is provided.}
    \label{fig:CalculationTI}
\end{figure}

Numerous works have been produced to evaluate and model the effect of the turbulence intensity on the power production of wind turbines \citep{Clifton2014,Bardal2017}. 
In this work, the impact of the turbulence intensity on the power curve is pragmatically calculated by assuming that short-term variations of the wind speed\footnote{ The typical maximum of the spectral density  of  the  wind  speed  has  its  maximum in  the  frequency  range  of  about  1/100 ~Hz, while  even  large  wind  turbines  can  accelerate  and  decelerate  the  rotor  within  only  a  few  seconds  (frequency  of  response  higher than 1/10~Hz). \citep{Albers2010}} follow a Gaussian distribution with mean $U=\mu (u)$ and standard deviation $U\cdot TI$ (see e.g. \citep{Albers2010}). 
With  this assumption the effect of the turbulence intensity on the power curve for a wind speed U can be considered by making a convolution between the original power curve and a Gaussian Kernel of mean $U$ and standard deviation $U\cdot TI$ and taking the resulting power for the wind speed U. 
This calculation is illustrated in \autoref{fig:CalculationTI}. 

In \autoref{fig:CalculationTI}, the upper, middle and lower plots represent respectively the original power curve, the different Kernels and the final power curve. 
The modified power value is the weighted average of power values at wind speeds between 9 and 13 where the weights are the blue kernel of the middle plot. 
The vertical light grey lines represent the weighted average. 
The same procedure is iterated for each wind speed with the different kernels represented in the middle plot. 

\begin{figure}[h!]
 \centering
    \includegraphics[width=0.8\textwidth]{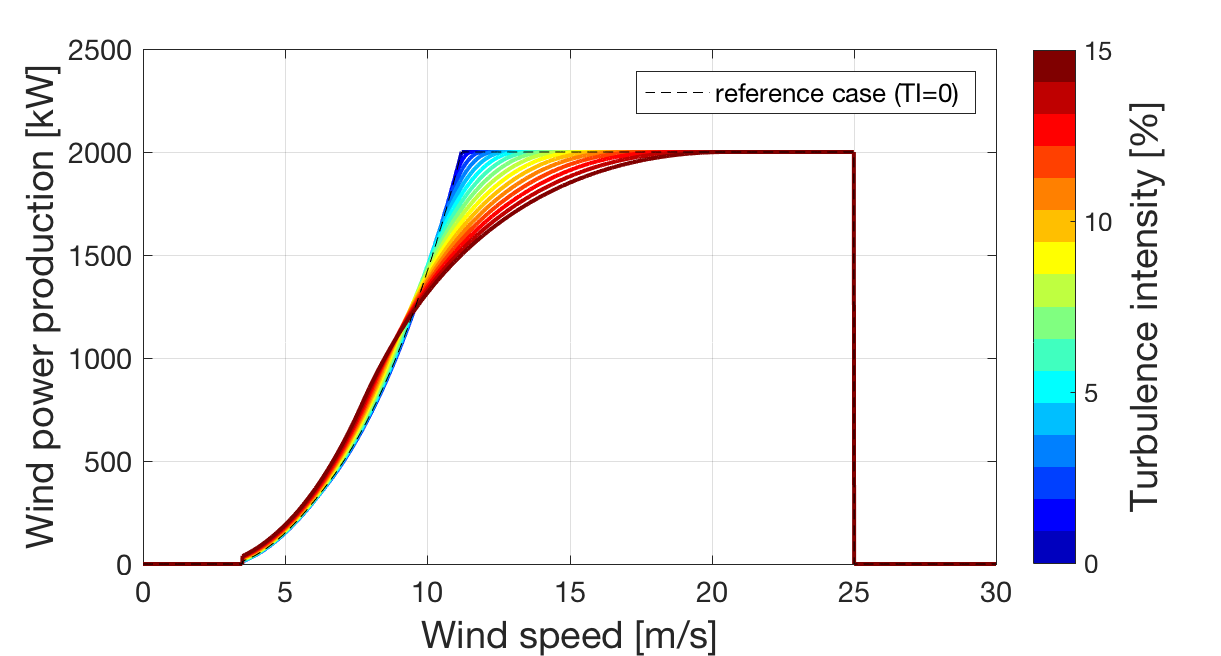}
    \caption{Illustration of the effect of the turbulence intensity on a power curve}
    \label{fig:EffectTI}
\end{figure}

The effect of the turbulence intensity on a power curve is illustrated in \autoref{fig:EffectTI} for different values of the turbulence intensity between 0 and 15\%, using the power curve of a 2-MW wind turbine with a rotor diameter of 80~m. 
This example shows clearly that the effect of the turbulence intensity can be significant, especially around the nominal wind speed. 
This parameter is therefore of paramount importance for the estimation of the power curve in real condition. 
It will be taken into consideration in the comparison of the model output with manufacturers power curves in \autoref{sec:Validation}.

As can be seen in \autoref{fig:EffectTI}, the turbulence intensity has no effect on the sudden power decrease as the wind speed exceeds its cut-off value.
It was indeed decided not to apply the smoothing effect of the turbulence intensity in this region since the cut-off is not activated based on high frequency wind speed but based on a longer time average.
In addition, an hysteresis implemented for the restart of the wind turbine as the wind speed decreases below the cut-off value hinders using the kernel convolution approach for the calculation of the TI effect on the power production. 

\bigbreak

\textbf{The effects of the air density on the power curve}

\begin{figure}[h!]
 \centering
    \includegraphics[width=0.8\textwidth]{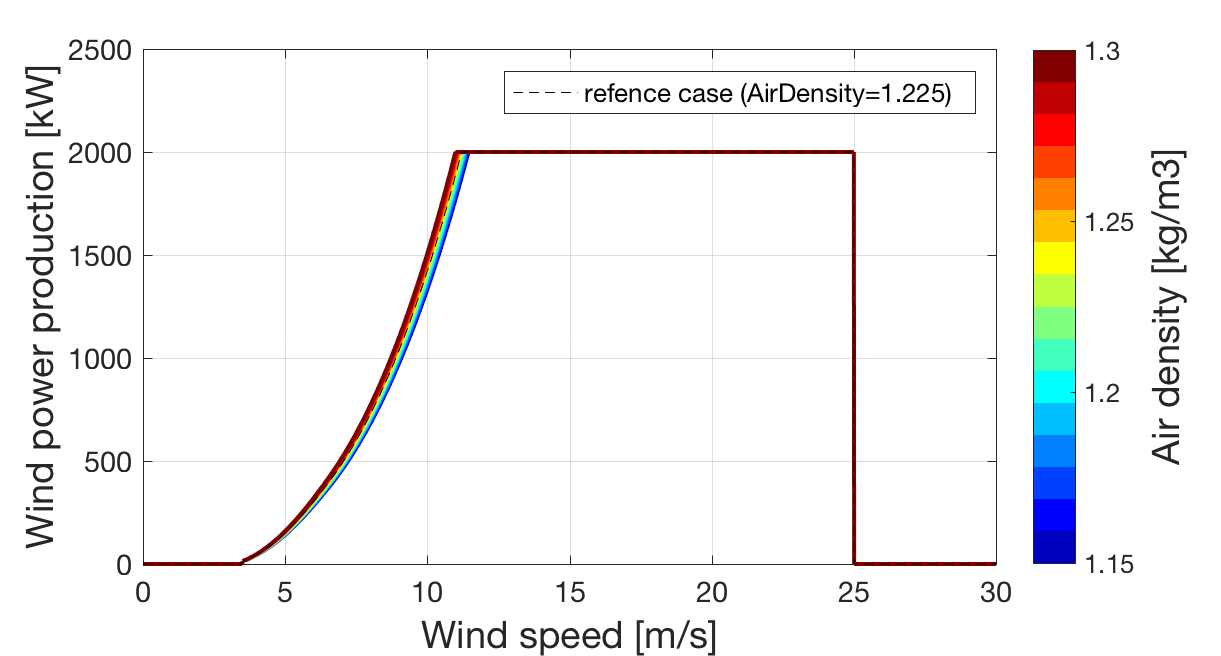}
    \caption{Illustration of the effect of the air density on a power curve}
    \label{fig:EffectAirdensity}
\end{figure}

With the approach proposed in this work, the consideration of air density on the power curve is explicit and straightforward, as is illustrated for values varying between 1.15 and 1.3 $kg/m^{3}$ in \autoref{fig:EffectAirdensity}. 
The reference value for the air density is set to 1.225 $kg/m^{3}$, which lies in the middle of the variation interval.
It can be observed in \autoref{fig:EffectAirdensity} that the impact of varying air density on the power curve is much lower than the effect of the turbulence intensity. 
Yet, it impacts the power curve across the whole range of Region II, where the frequency of occurrence is generally high and a careful consideration of this external factor should therefore be made.

\textbf{The effects of the wind shear and the wind veer on the power curve}

\begin{figure}[h!]
 \centering
    \includegraphics[width=0.9\textwidth]{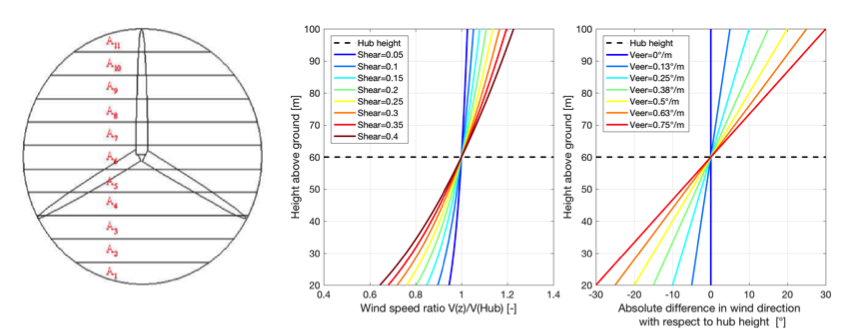}
    \caption{ Illustration of wind shear and wind veer across the wind turbine rotor. The left panel illustrates a turbine rotor divided into horizontal bands, corresponding to those in Eq.~\autoref{eq:9}. The middle and right panels illustrate the variation in wind speed and direction with height. }
    \label{fig:DefinitionSheerVeer}
\end{figure}

Wind speed is not uniform across the wind turbine's rotor plane, as it increases with height through the atmospheric boundary layer. 
The vertical wind profile can be described in several ways \citep{SCHALLENBERGRODRIGUEZ2013272}, such as the logarithmic profile which depends on the roughness length, friction velocity and stability parameter. 
In many applications, the simpler power law model is used which relates the ratio of the wind speeds at two heights with the power of the ratio of the two heights:

\begin{equation}
\label{eq:7}
       u(z)=u(z_{hub})\cdot \left( \frac{z}{z_{hub}}\right)^{\alpha}
\end{equation}
\bigbreak
Where $u(z_{hub})$ is the wind speed at hub height, $z_{hub}$, and $u(z)$ is the wind speed at height $z$. 
In this equation, $\alpha$ is the Hellman or shear coefficient, which quantifies the wind shear and varies typically between 0 and 0.4. 
The effect of the wind shear on the vertical profile of the wind speed in the region of a rotor area is illustrated in the middle panel of \autoref{fig:DefinitionSheerVeer}. 
In this example, a hub height of 60 meters and a rotor diameter of 80 meters have been assumed.

Wind veer is defined as the change in wind direction as a function of height. It has been shown that wind veer does exist in typical wind situations \citep{Ivanell2010}. 
To consider wind veer, we assume that the change in wind direction is zero at hub height and varies linearly with height according to:

\begin{equation}
\label{eq:8}
       \Delta \varphi (z)=v\cdot(z-z_{hub})
\end{equation}
\bigbreak
In the above equation, the parameter $v$ quantifies the evolution of the difference in wind direction $\Delta \varphi (z)$ as a function of the height difference $(z-z_{hub})$.
Based on the statistical analysis of \citet{Ivanell2010}, we assume that this parameter can vary between 0 and 0.75°/m. 
The wind veer is illustrated in the right plot of \autoref{fig:DefinitionSheerVeer}. 

To evaluate the impact of wind shear and veer on the power curve, we follow the approach that is recommended in a revision of the IEC standard \citep{IEC61400-12-1_Ed2}, which consists in replacing the wind speed at hub height by a rotor equivalent wind speed $U_{eq}$, which is defined as:

\begin{equation}
\label{eq:9}
       U_{eq}=\sqrt[3]{\sum_{i} \left(\frac{A_{i}}{A}\right)\cdot \left ( U_{i} \cdot cos \left ( \Delta \varphi_{i}\right)\right)^{3} }
\end{equation}
\bigbreak

The coefficients $A_{i}$ correspond to the area of elementary horizontal bands of the rotor area as illustrated in the left panel of \autoref{fig:DefinitionSheerVeer}.
$U_{i}$ and $\Delta \varphi_{i}$ corresponds respectively to the wind speed and variation of the wind direction with respect to that at hub height in the $i^{th}$ horizontal band.

 \begin{figure}[h!]
 \centering
    \includegraphics[width=0.99\textwidth]{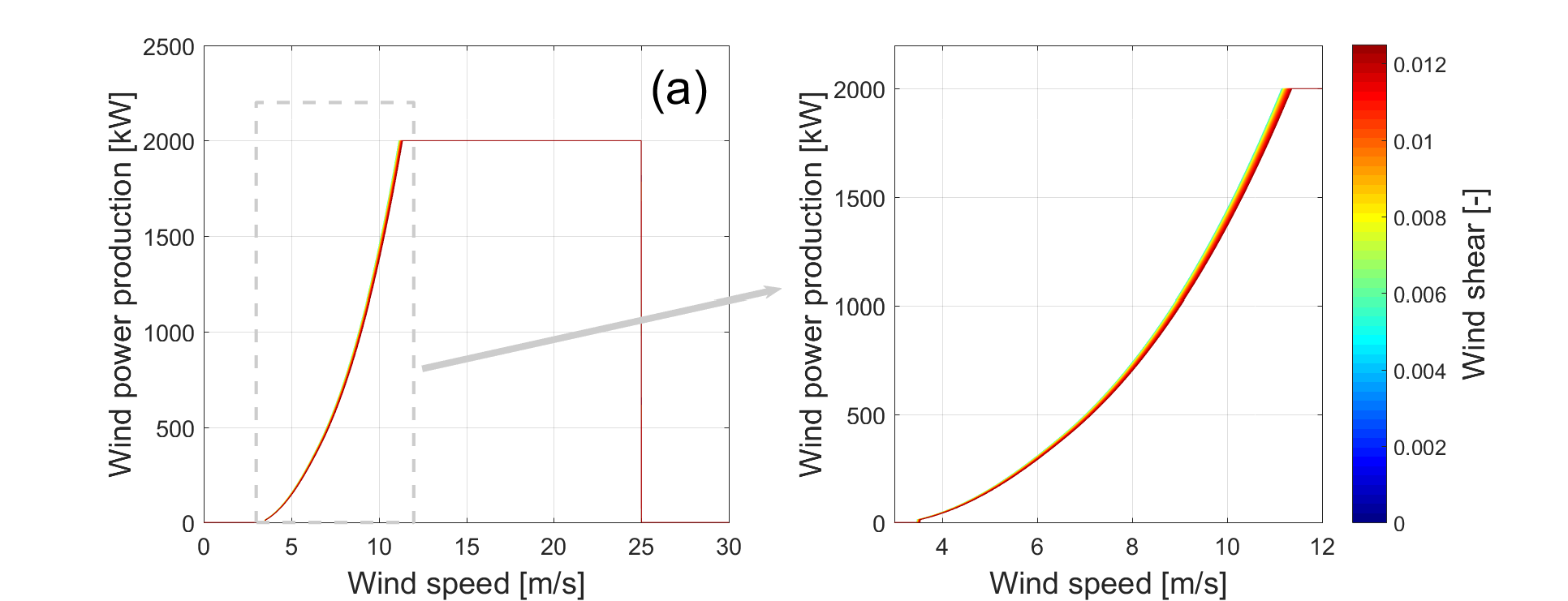}
    \includegraphics[width=0.99\textwidth]{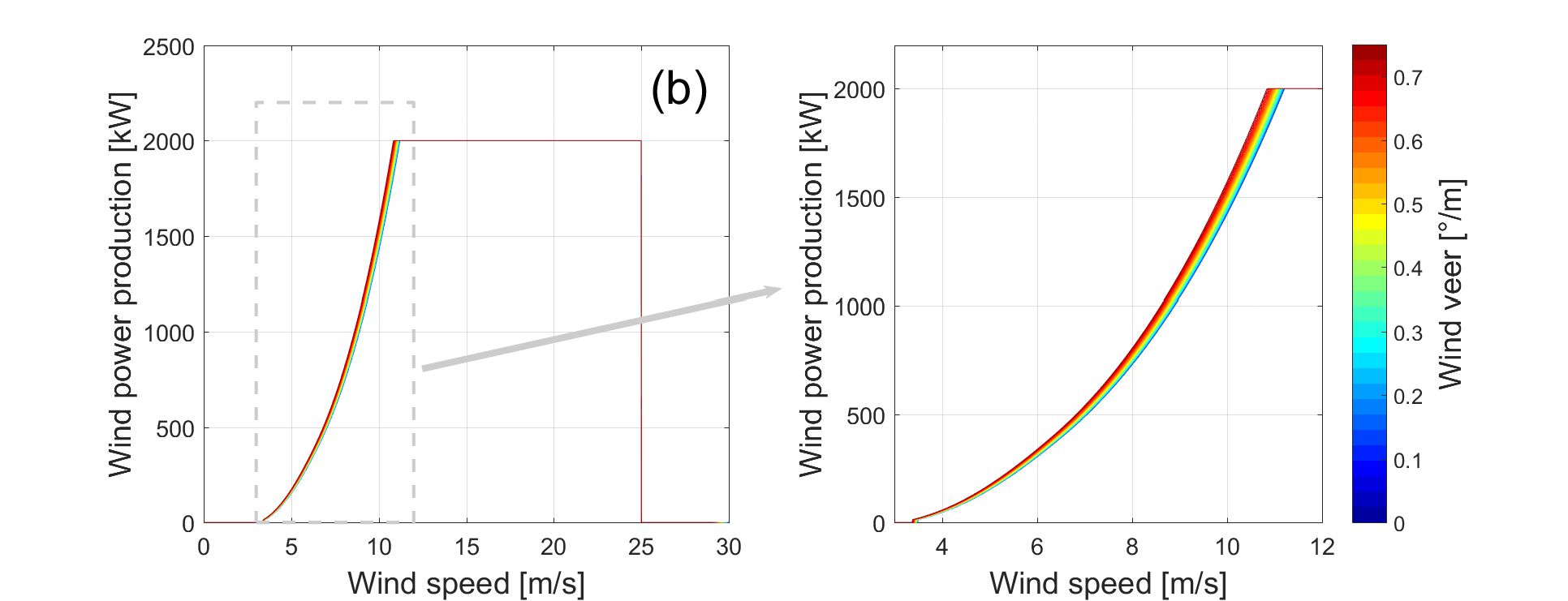}
    \caption{Influence of the wind shear (a) and the wind veer (b) on the power curve of a typical 2-MW wind turbine (80-meter rotor diameter and hub-height of 60 meters)}
    \label{fig:EffectSheerVeer}
\end{figure}

The influences of the wind shear and veer on the power curve of a 2-MW wind turbine with a rotor diameter of 80 meter and a hub height of 60 meter are represented in \autoref{fig:EffectSheerVeer}. 
The variations in the power curve illustrated in these two plots were obtained by replacing the wind speed at hub height by the equivalent wind speed calculated with Eq.~\autoref{eq:9}. 

In the upper plot of \autoref{fig:EffectSheerVeer}, the impact of the wind shear is barely visible, which is in agreement with the work of \citet{Wagner2010}.
The limited effect of the wind shear on the power production can be explained by two factors. 
Firstly, larger values of the cubic wind speed above hub height are balanced by lower values below hub height.
It should be however noted that this balancing effect becomes limited as the vertical distance to the hub height increases.
Secondly, the impact of wind speed values far from the hub height are limited by the area of the horizontal rotor band, which decreases with increasing distance to the hub height.

In the lower plot of \autoref{fig:EffectSheerVeer}, it can be observed that the impact of the wind veer is small yet greater than that of the wind shear. 
Indeed, the effect of wind veer is larger that that of the wind shear because the effective wind speed decreases above and under the hub so that there is no balancing effect.
Yet, the weighting resulting from the horizontal bands of rotor area limits the effect of the vertical change in wind direction on REWS.

We can also observe in  \autoref{fig:EffectSheerVeer} that the wind shear and veer are not impacting the power curve in the cut-off region; we consider indeed that the cut-off is activated based on the measurements of wind speed at hub height.

The wind shear and veer are both integrated in the codes accompanying this paper. It necessitates information on the hub height as additional parameter for the generation of power curve.

\subsection{Summary of the modelling approach}\label{sec:SummaryModel}
\begin{figure}[h!]
 \centering
    \includegraphics[width=\textwidth]{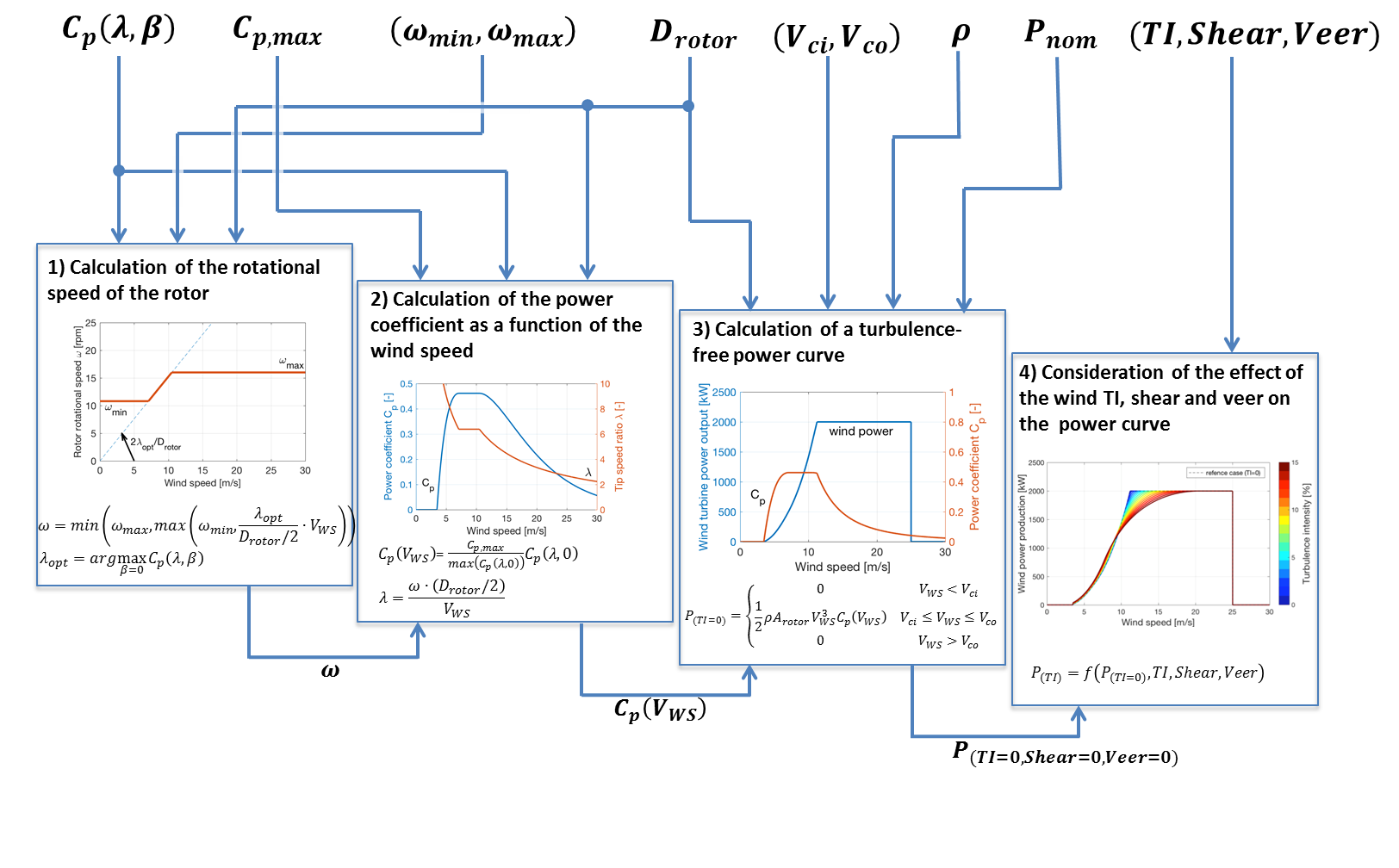}
    \caption{ Flow chart representing the main computation steps for the estimation of a power curve from characteristics of a wind turbine}
    \label{fig:SummaryCalcSteps}
\end{figure}

The main computational steps described in the previous sections are summarised in \autoref{fig:SummaryCalcSteps}. 
In the first step, the rotor speed is evaluated as a function of the wind speed. 
This is achieved using the optimal tip-speed ratio evaluated with Eq. \autoref{eq:4} and the relationship between wind speed and tip-speed ratio given by Eq.~\autoref{eq:3}, respecting the operational range of the rotor speed $[\omega_{min}; \omega_{max}]$. 
In the second step, the evolution of the power coefficient with wind speed is evaluated using the rotor speed $\omega$ and a power function $C_{p}(\lambda,\beta)$.
The analytical expression given in Eq.~\autoref{eq:2} is used in this work, but other expressions can be implemented instead. 
In order to differentiate the form of the $C_{p}(\lambda,\beta)$ to its maximal value, the function $C_{p}(\lambda,\beta)$ is scaled so that its maximal value is the newly introduced parameter $C_{p,max}$. 
The power output of the wind turbine is evaluated in the third step using Eq.~\autoref{eq:1}.  This curve is scaled by the nominal power of the turbine, and then the cut-in and cut-off wind speeds are applied. 
The fourth and final step introduces the effect of the turbulence intensity on the power curve.
This is the only external effect that is considered explicitly in our model, since other effects can be applied by evaluating a rotor equivalent power curve \citep{IEC61400-12-1_Ed2}.

The input parameters of the model are listed in the upper line of \autoref{fig:SummaryCalcSteps}. 
These input parameters can be gathered in three groups of different natures. 
Firstly, two parameters $C_{p}(\lambda,\beta)$ and $C_{p,max}$ are related to the aerodynamic efficiency of the blades. 
The second group of parameters ($TI$ and $\rho_{air}$) are related to external conditions. 
Finally, the last group of parameter includes 6 design characteristics of wind turbines that can in most cases be found in manufacturer's product sheets. 

As argued in \citep{PFENNINGER2017211}, energy models should be made open to improve the quality of science and aid the productivity of other researchers. We therefore implement the model outlined in \autoref{fig:SummaryCalcSteps} in three widely-used programming languages, Python, R and MATLAB. The power curve for an arbitrary wind turbine can be generated by simply specifying the rotor diameter and nominal power output.  Sensible defaults are given for all other parameters, or they can be customised as desired.  The model code is available from Github \url{https://github.com/YvesMSaintDrenan/WT_PowerCurveModel}.

\section{Analysis of the sensitivity of the power curve to the model parameters}\label{sec:sensitivity}
A sensitivity analysis was performed to assess the relative importance of the different parameters of the model. 
A reference set of parameters was determined and the sensitivity of the power curve to each parameter was evaluated by varying each parameter individually across a typical range. 
The set of reference parameters and their variation interval are given in \autoref{T:RefValues4SensAn}.

\begin{figure}[h!]
    \includegraphics[width=0.49\textwidth]{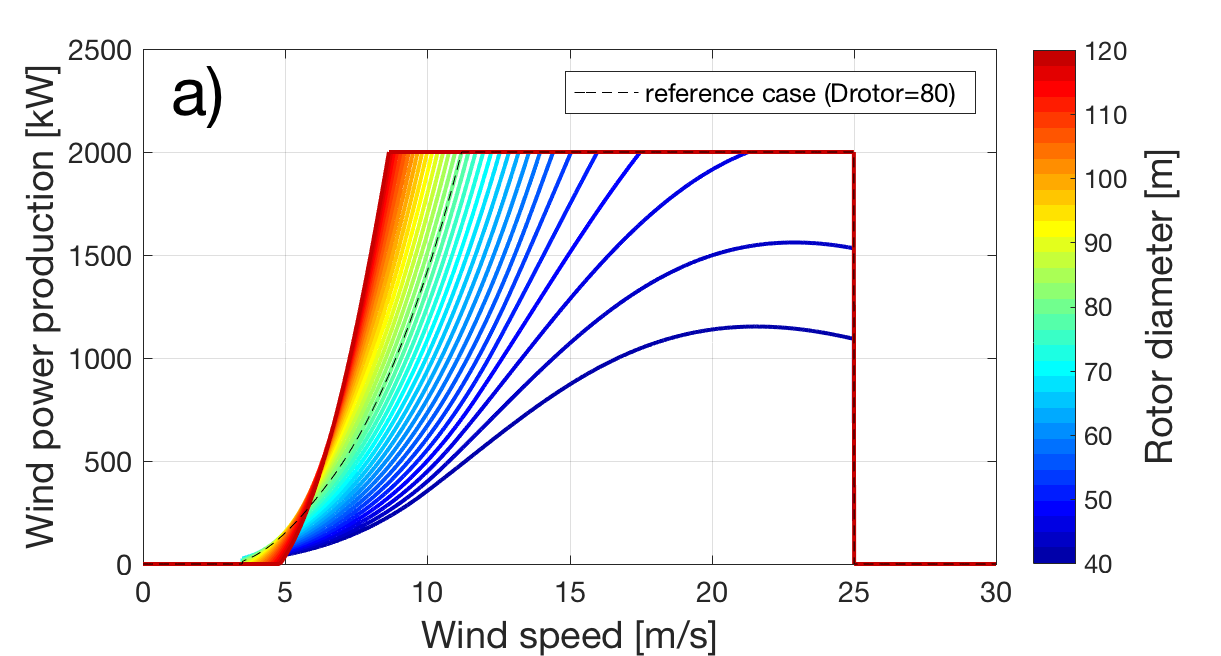}
    \includegraphics[width=0.49\textwidth]{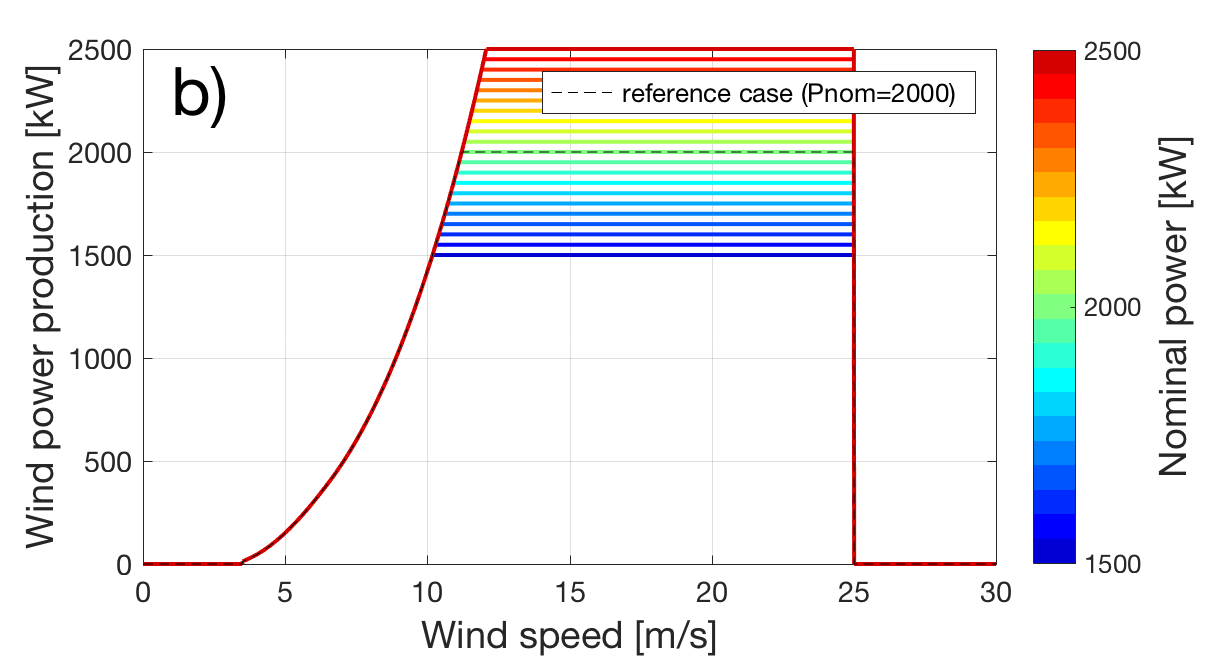}
    \includegraphics[width=0.49\textwidth]{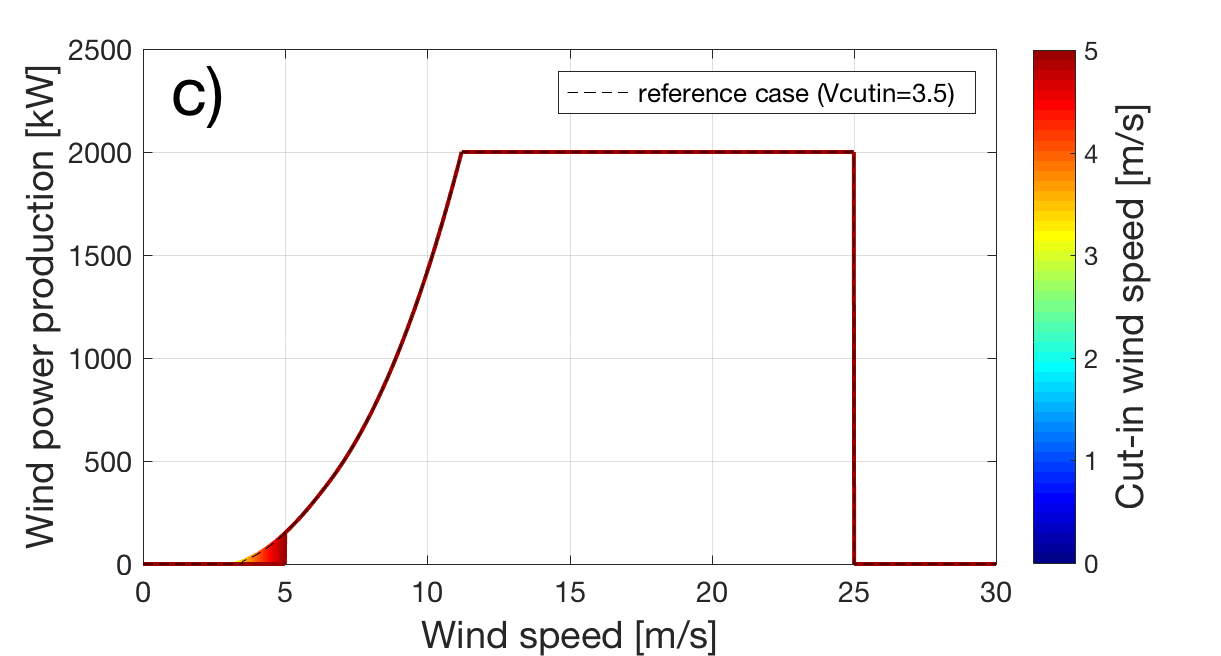}
    \includegraphics[width=0.49\textwidth]{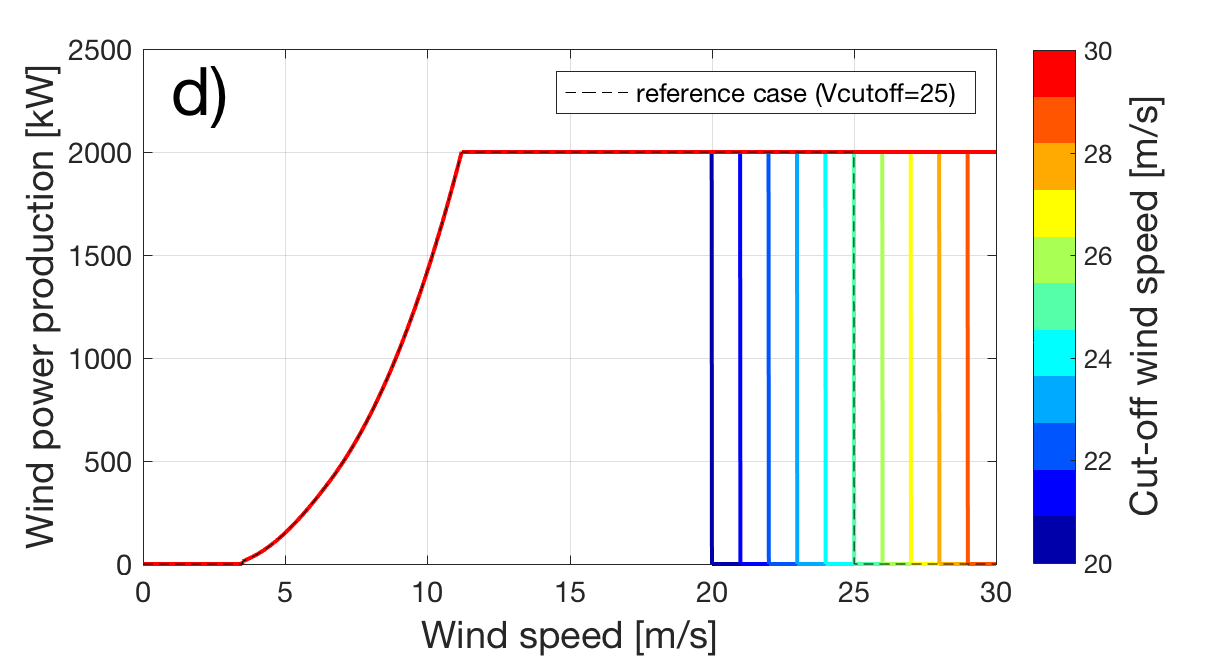}
    \includegraphics[width=0.49\textwidth]{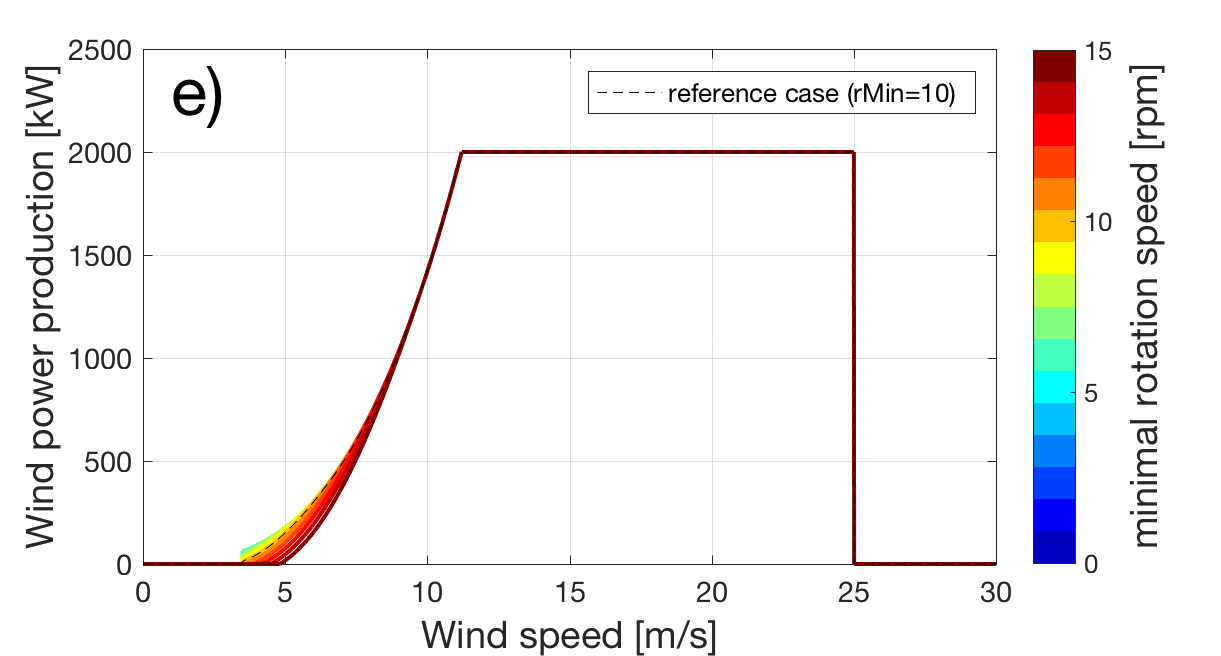}
    \includegraphics[width=0.49\textwidth]{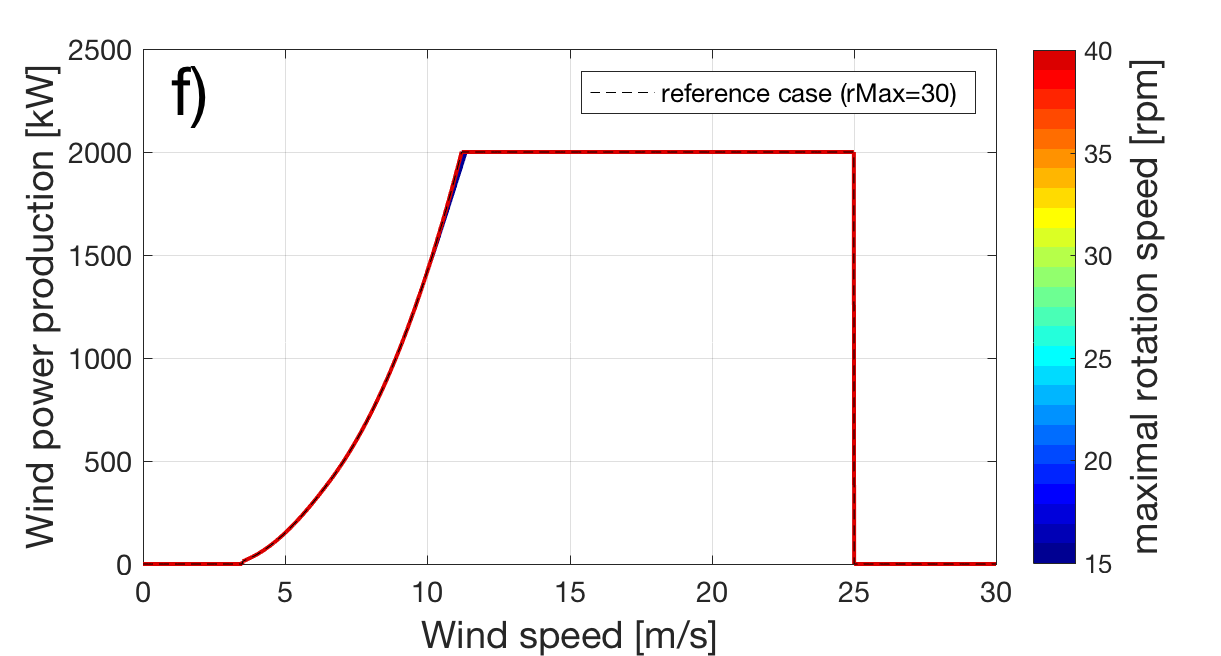}
    \includegraphics[width=0.49\textwidth]{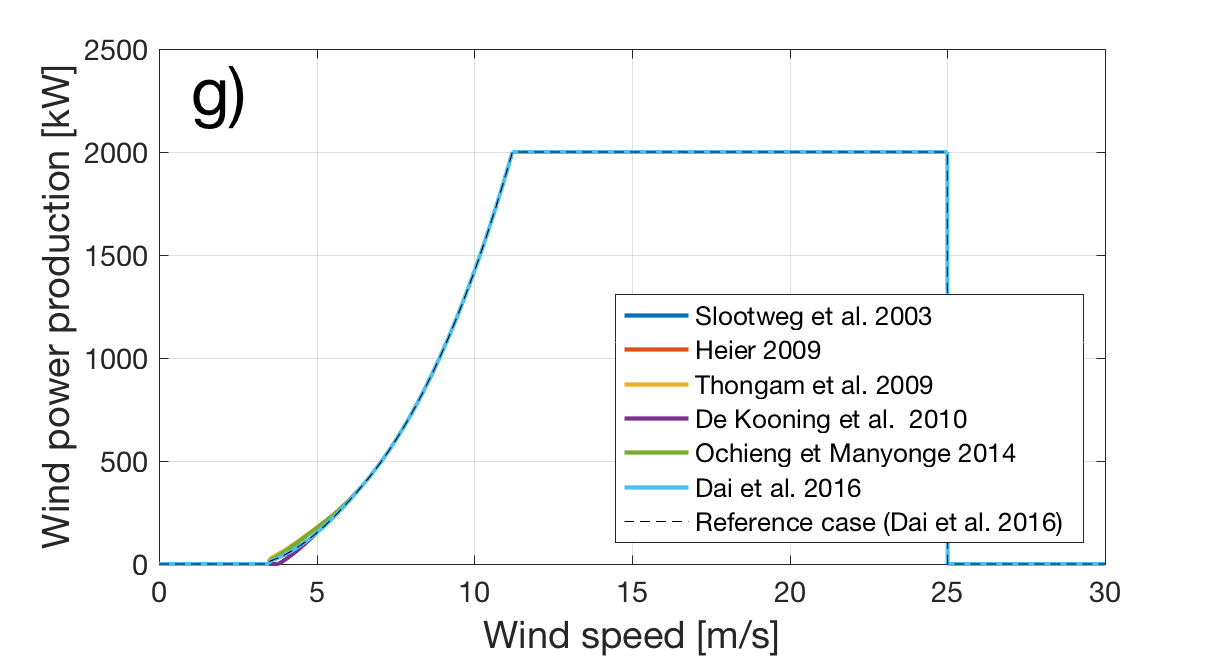}
    \includegraphics[width=0.49\textwidth]{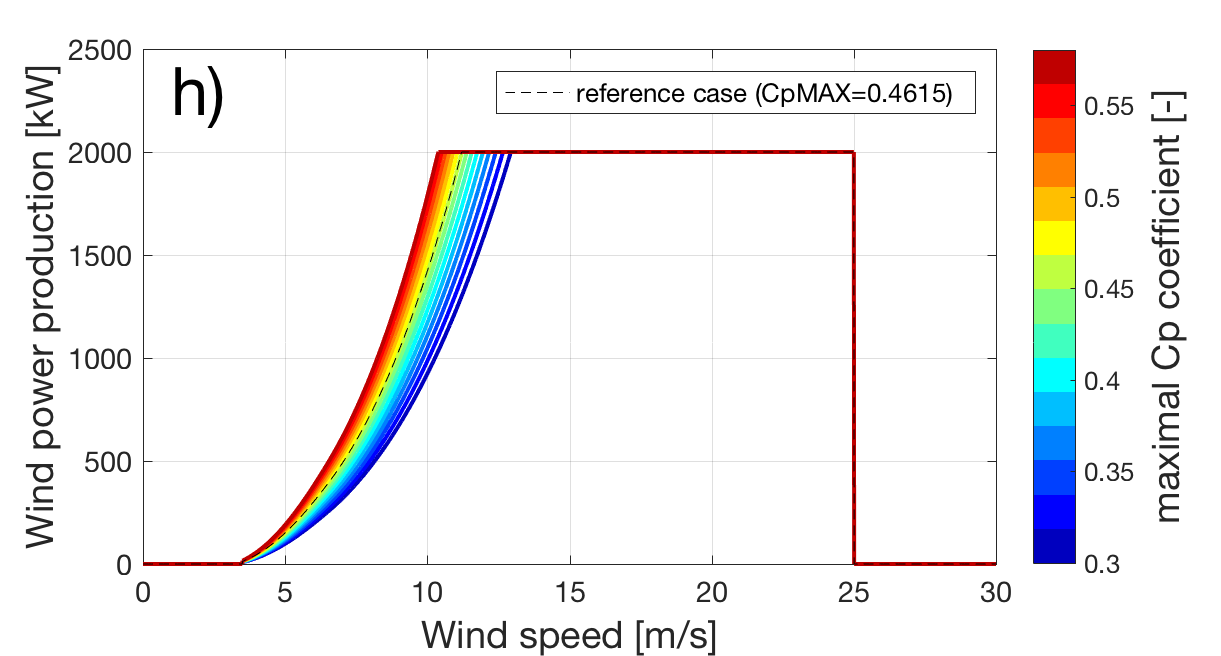}
    \caption{Influence of the different parameters on the power curve. In each panel one input parameter is varied across a typical range, as given in the panel's legend. }
    \label{fig:ResultsSENSAN}
\end{figure}

The present sensitivity analysis is limited to a univariate analysis: one parameter is varied at the time. 
The sensitivities of the parameters are not quantified as in the Morris screening method \citep{Morris1991} or generalised sensitivity  analysis \citep{Sobol1993} but only qualitatively assessed. 
For this purpose, the sensitivities are visually represented by lines of different colours for the different values of the varied parameters in the plots of \autoref{fig:ResultsSENSAN}.

\begin{table}[!h]
\begin{center}
\begin{tabular}{| c | c | c |}
\hline
\textbf{Parameters} & \textbf{Reference value} & \textbf{Variation interval} \\\hline
Rotor diameter & 80 m & 40 - 120 m\\\hline
Nominal power & 2000 kW& 1500 - 2500 kW\\\hline
Cut-in wind speed & 3.5 m/s& 0 - 5 m/s\\\hline
Cut-out wind speed & 25 m/s & 20 - 30 m/s\\\hline
Minimal rotation speed & 10 rpm & 0 - 15 rpm\\\hline
Maximal rotation speed & 30 rpm & 15 - 40 rpm\\\hline
Maximum $C_p$ value & 0.4615 & 0.3 - 0.59 \\\hline
& & \citep{Slootweg2003}\\
& & \citep{Heier2014}\\
$C_p$ parameterisation & \citep{Dai2016} & \citep{Thongam2009}\\
& & \citep{DeKooning2013}\\
& & \citep{Ochieng2014}\\
& & \citep{Dai2016}\\ \hline
\end{tabular}
\end{center}
\caption{ Reference values and variation intervals of the different parameters considered in the sensitivity analysis}
\label{T:RefValues4SensAn}
\end{table}

The sensitivity of the reference power curve to variations of the rotor area and the nominal power are displayed in the first row of \autoref{fig:ResultsSENSAN} (plots (a) and (b)). 
These two parameters yield the largest sensitivity to the output power and should thus be treated with the greatest caution. 
However, their level of uncertainty is negligible as they both are design parameters and most manufacturers mention them directly in the name of the turbine (e.g. Vestas V80-2000, Enercon E82 E2/2.0MW, GE Haliade 150-6MW, Gamesa G114-2.0MW, Bonus B82/2300…).

It is interesting to note that for very small rotors, the power curves move away from a cubic increase and even decrease at high wind speed.
This is due to the rotational speed that increases with decreasing rotor area to reach the optimal TSR.
As soon as the rotational speed is bounded by the maximal rotational speed, increase of the wind speed brings about decrease of the power coefficient which can ultimately results in a decrease of the power (blue curves in \autoref{fig:ResultsSENSAN}-a). 

The effects of variations of the cut-in and cut-off wind speeds on the reference wind turbine are illustrated in the second row of \autoref{fig:ResultsSENSAN} (plots (c) and (d)). 
It can be observed that the cut-off wind speed has the largest impact on the power curve while the sensitivity of the power curve to the cut-in wind speed is moderate. 
Though, it should be noted that the frequency of occurrence of wind speed in the neighbourhood of the cut-in wind speed can be high while wind speed values close to the cut-off wind speed are much less frequent. 
The two wind speeds are therefore both important parameters for the estimation of the annual energy production of a wind turbine but they both are of lesser importance compared to other parameters such as the nominal power or the rotor area.

The sensitivities of the model output to the minimum and maximum rotation speeds are displayed in the third row of \autoref{fig:ResultsSENSAN}. 
It can be observed that the effect of the minimum rotational speed is limited to a wind speed interval of 3-9 m/s while that of the maximum rotational speed can be observed for wind speed values close to the nominal wind speed. 
This is due to the fact that the rotation speed is unconstrained between these two intervals. The effect of the maximum rotor speed is hardly visible. 
This is due to two reasons: firstly, the wind speed corresponding to the maximum rotation speed is very close to the nominal speed and, secondly, the decrease of the $C_{p}$ value with the wind speed at maximal rotation speed is relatively small. 
The sensitivity of the model to the minimum rotation speed is more pronounced and should accordingly be carefully chosen since the frequency of occurrence of the wind speed in the interval 3-9 m/s is high.

In the last row of plots, the sensitivity of the model output to the scaled $C_{p}$ model and the maximal value of the power coefficient are displayed. 
We rescaled the shape of the $C_{p}$ model to the magnitude of $C_{p}$ values given by the model and proposed to scale the model output using the new parameters $C_{p,max}$, which corresponds to the maximal power coefficient given by the model. 
This is very instructive since as can be observed in the plots (e) and (f) of \autoref{fig:ResultsSENSAN}, the power curve is not sensitive to the choice of the scaled parameterisation but the effect of variations of the parameter $C_{p,max}$ on the power curve is significant. 
This shows that the choice of the $C_{p}$ model is not critical but the choice of an accurate value for $C_{p,max}$ is decisive for an accurate calculation of the wind power production.

\section{Statistical analysis of the most sensitive model input parameters}\label{sec:StatisticalAnalysis}

While parameters such as the nominal power and the rotor area are readily available for each turbine, this is not the case for other parameters such as the maximum power coefficient or the rotor minimal and maximal speeds. 
In order to address such situations, a statistical analysis of the model parameters was performed, which can be used as guidance to the choice of unknown parameters. 
For this, we used the database of wind turbines and power curves provided by \citet{thewindpower}, which includes extended data on the main turbine characteristics. 
As of May 2019, this commercial database contains about 780 turbines models.

\subsection{Maximum value of the power coefficient $C_{p,max}$ }

\begin{figure}[h!]
    \centering
    \includegraphics[width=0.9\textwidth]{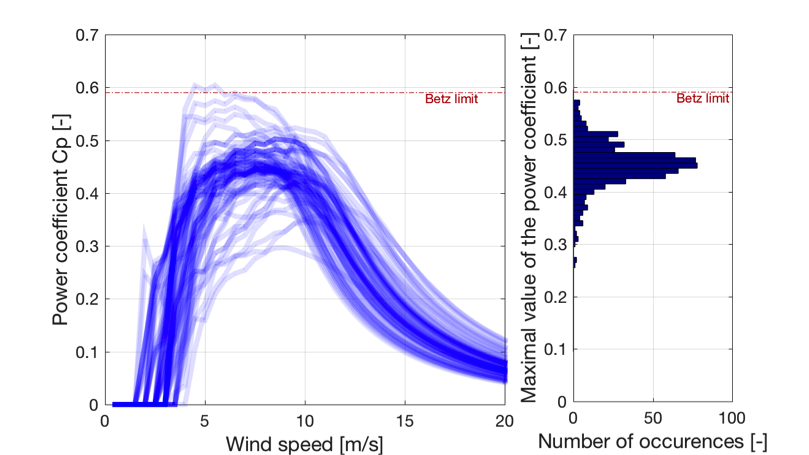}
    \caption{Left: Power coefficient as a function of the wind speed for the power curves available in \citet{thewindpower} dataset. Right: distribution of the maximal power coefficient evaluated for all available power curves. }
    \label{fig:StatisticsCpMax}
\end{figure}

The value of $C_{p,max}$ has been evaluated for 600 wind turbines using the power curve and characteristics of the turbine by inverting Eq.~\autoref{eq:1} and selecting the maximum value. 
This process is illustrated in the two plots of \autoref{fig:StatisticsCpMax}. 
The power coefficient is plotted as a function of wind speed for all turbines in the left plot, and a histogram of the maximum values for each turbine is shown in the right plot. 
It can first be noticed that there are some potentially corrupted values of $C_{p}$; these will be discussed in the next section. 
The most frequent value is 0.44 and 80 \% of the values are between 0.4 and 0.5. 
A dependency of this parameter on further characteristics such as e.g. the size of the turbine can be expected but no clear dependencies could be identified with the available data. 
Based on this short analysis, we therefore recommend using a value of 0.44 when this information is not available. 

\subsection{Cut-in and cut-off wind speeds}

\begin{figure}[h!]
    \centering
    \includegraphics[width=0.7\textwidth]{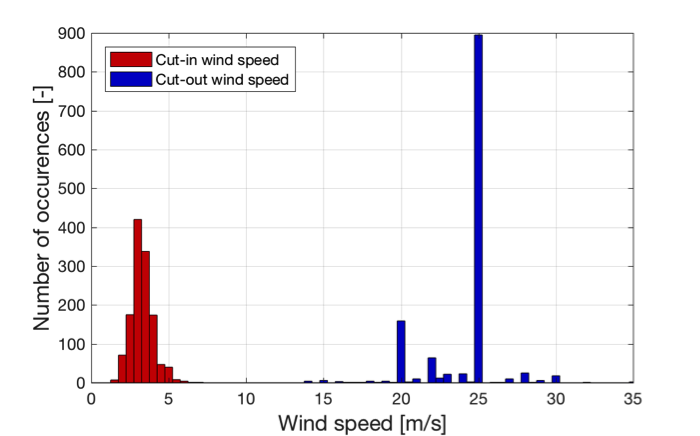}
    \caption{Distribution of the cut-off wind speed from the European dataset of \citet{thewindpower}}
    \label{fig:StatisticsCICO}
\end{figure}

The distributions of the cut-in and cut-off wind speeds of the wind turbine information contained in \citet{thewindpower} dataset are displayed in \autoref{fig:StatisticsCICO}. 
It can be observed that the cut-in wind speed are between 1 and 5 m/s. 
Most values are distributed around 3 m/s with 90 \% of the values between 2 and 4 m/s. 
The cut-off wind speed are between 15 and 30 m/s. 
The most frequent values are 20 and 25 m/s, with a share of all wind turbines of respectively 12 \% and 70 \%. 
When information on the cut-in and/or cut off wind speeds is unavailable, values of respectively 3 and 25 m/s can be recommended based on the present analysis.

\subsection{Minimal and maximal rotational speed}

\begin{figure}[h!]
    \centering
    \includegraphics[width=0.8\textwidth]{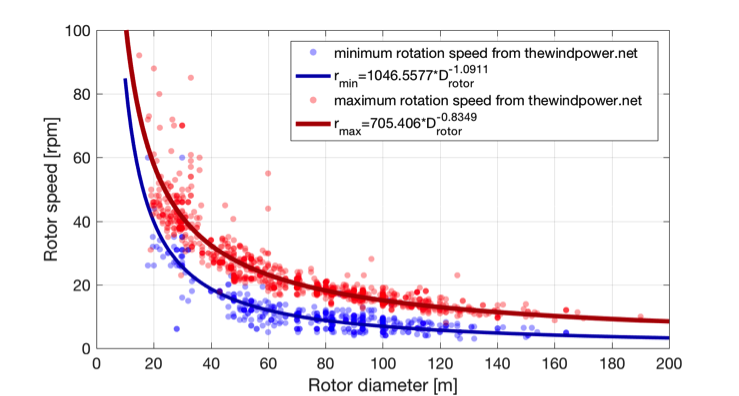}
    \caption{Distribution of the minimal and maximal rotational speed from the European dataset of \citet{thewindpower}}
    \label{fig:StatisticsRPMminmax}
\end{figure}

As can be observed in \autoref{fig:StatisticsRPMminmax}, the minimum and maximum rotational speeds exhibit a strong dependency on the rotor diameter.
The same statistical analysis as those presented in the two previous subsections could therefore not be carried out. 
Instead, we adopted a similar approach as that described in \citep{LaraGarcia2013} and fitted this dependency using an exponential function, which gives the two following expressions for the minimal and maximal rotation speed as a function of the rotor diameter:

\begin{equation}
\label{eq:10}
       \omega_{min}=a\cdot D_{rotor}^{b} \: with \: \begin{cases} a=1046.558 \\ b=-1.0911\end{cases}
\end{equation}
\begin{equation}
\label{eq:11}
       \omega_{max}=c\cdot D_{rotor}^{d} \: with \: \begin{cases} c=705.406 \\ b=-0.8349\end{cases}
\end{equation}
\bigbreak

The minimum and maximum rotation speed can be estimated with Eq.~\autoref{eq:10} and Eq.~\autoref{eq:11} when information on these characteristics is missing.

\section{Validation of the parametric power curve model}\label{sec:Validation}

\begin{figure}[h!]
    \includegraphics[width=\textwidth]{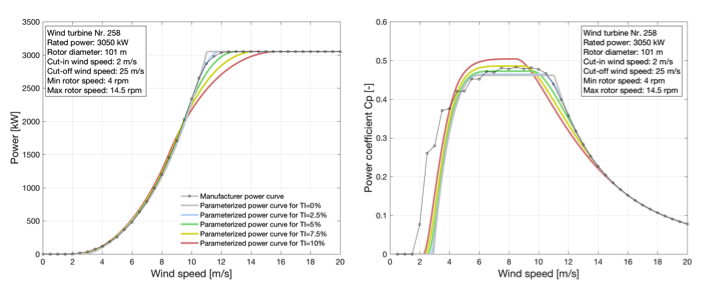}
    \includegraphics[width=\textwidth]{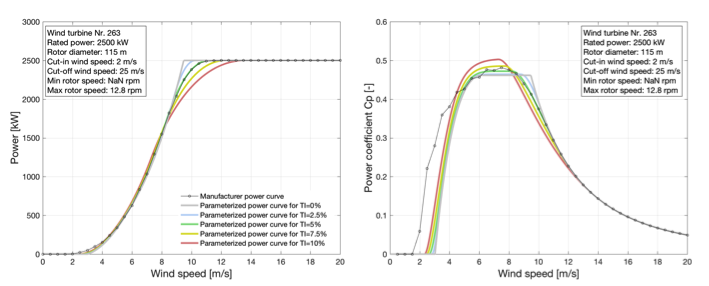}
    \includegraphics[width=\textwidth]{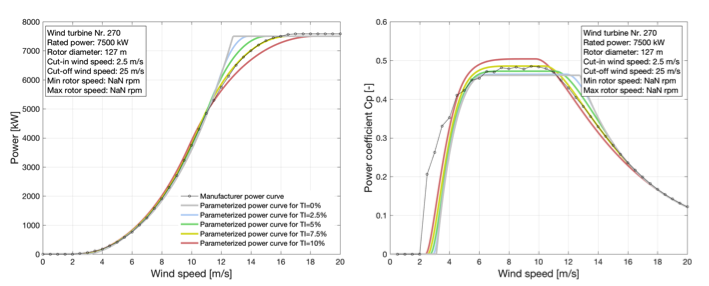}
    \caption{Comparison of the model output with the power curves of three wind turbines which are modelled with high quality.  The three rows show the wind turbines 258, 263 and 270 of \citet{thewindpower}}
    \label{fig:Validation_11}
\end{figure}

A validation of the model introduced in \autoref{sec:Methodology} has been conducted using the power curve of 91 wind turbines with a nominal power greater than 1~MW as provided by their manufacturers and available in the \citet{thewindpower} dataset. 
For this validation, the power curve model has been run with specific information on the nominal power and rotor area, while other model inputs are set to the reference values described in \autoref{sec:StatisticalAnalysis} and the air density is set to 1.225 kg/m\textsuperscript{3}. 
As the level of turbulence intensity corresponding to each power curve is unknown, they are compared to model outputs obtained with values of the turbulence intensity ranging between 0 and 10 \%. 
As a consequence, a quantitative validation could not be conducted, as this unknown parameter would have to be optimised in the model (which would instead be calibration).  Rather, a qualitative validation is performed through a visual comparison of the model output to the database of power curves, whose main outcomes are described in this section. 
In \autoref{fig:Validation_11}, the result of the validation is given for three wind turbines which show a close correspondence between model output and power curve from the database. While this does not completely exclude the possibility of a systematic modelling error, the degree of similarity across the plots suggests the model can synthesise realistic power curves for a variety of wind turbines.

It can be observed that the best matches between model output and turbine data are obtained for different values of the turbulence intensity: 2.5, 5 and 7.5\% for the wind turbine 258, 263 and 270. 
As mentioned above, information on the turbulence intensity is not available in the database, which hinders a proper validation of the model and represents a non-negligible source of uncertainty for power estimation made with these power curves taken from manufacturers and other sources. 
This highlights an advantage of the model we present, because it allows the value of the turbulence intensity to be controlled, or varied. 
This also applies to the effect of air density, allowing exploration of how the power curve evolves with the altitude, temperature and season.

\begin{figure}[h!]
    \includegraphics[width=\textwidth]{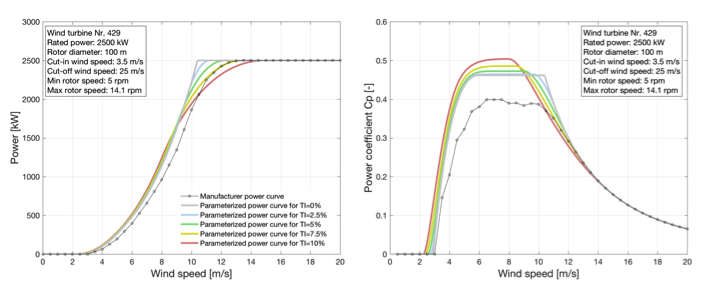}
    \includegraphics[width=\textwidth]{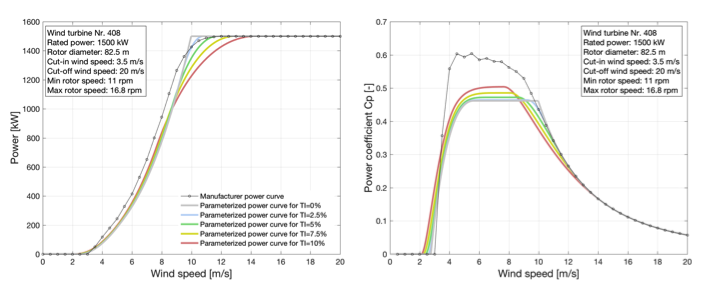}
    \caption{ Comparison of the model output with the power curves of two wind turbines which are modelled with lower quality.  The two rows show the wind turbines 429 and 408 of \citet{thewindpower}}
    \label{fig:Validation_12}
\end{figure}

Examples of wind turbines found to have the highest difference between model and turbine power curves in the validation are given in \autoref{fig:Validation_12}. 
The two plots on the right side of \autoref{fig:Validation_12} show that this difference stems from a mismatch between the maximal value of the power coefficient assumed in our model and the actual value of a wind turbine. In one case, for turbine 408 this is possibly due to an error in the measured power curve, which actually exceeds the Betz limit.

Such values are known to result from power curves measured with shaded anemometer \citep{Shin2019}. 
This observation highlights the need for careful screening for data quality when using power curves from manufacturers and databases.

\begin{figure}[h!]
    \includegraphics[width=\textwidth]{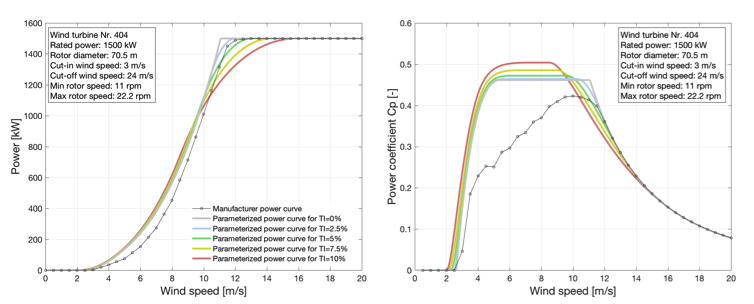}
    \caption{ Comparison of the model output with the power curve of wind turbine 404  of \citet{thewindpower}, which has an unknown divergence}
    \label{fig:Validation_13}
\end{figure}

In the examples presented so far the shape of the power coefficient function matches with that from the wind turbine database.  However, turbine 404 in \autoref{fig:Validation_13} exhibits a different shape which does not conform to the majority of other wind turbines.
In this case, it is difficult to identify the reason for the observed difference (modelling error, correction of the effect of turbulence intensity, shaded wind measurements…) and further validation work would be needed to get a deeper insight in the performance of the model and possible sources uncertainty on the power curves.

\begin{figure}[h!]
    \includegraphics[width=\textwidth]{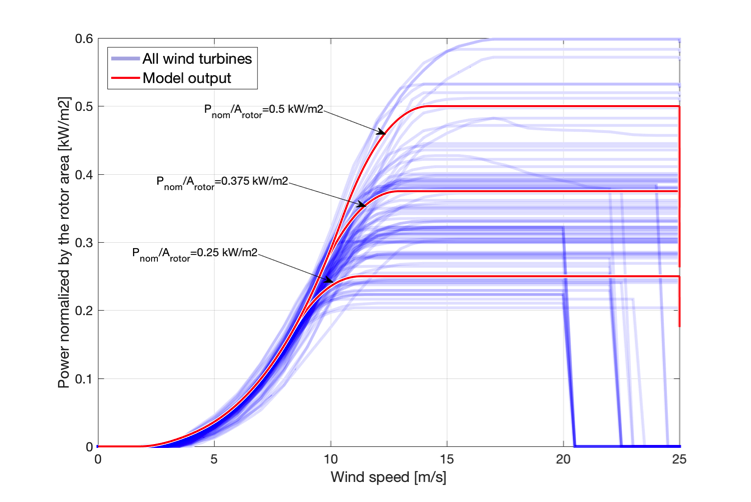}
    \caption{Comparison of all power curves contained in the database (blue curves) with the power output obtained for a power to rotor area ratio of 0.25, 0.375 and 0.5 kW/m2 (red curves)} (\underline{source:} \citet{thewindpower})
    \label{fig:Validation_14}
\end{figure}

The different examples presented above summarise most situations encountered in the validation work. 
In order to give an overview on the match between the model output and the database of power curves, all power curves have been represented by blue lines in \autoref{fig:Validation_14}. 
Since the error of the model principally occurs in region II, the power curves have been scaled by normalising them  by the rotor area so that all power curves are similar in that area. 
Indeed, it can be observed that many of them overlap for wind speed values between 3 and 10 m/s. 
The model outputs obtained with the standard parameters but with power to rotor area ratio of 0.25, 0.375 and 0.5 and a turbulence intensity of 5\% have also been represented in this plot.

This final analysis shows that even in cases with the largest errors, such as those presented in \autoref{fig:Validation_12} and \autoref{fig:Validation_13}, the model is able to reliably reproduce the behaviour of the majority of power curves seen across the wind industry.

\section{Conclusion}\label{sec:Conclusion}

We propose here an approach to estimate the power curve of a wind turbine from its main characteristics. 
The present work has aggregated existing knowledge on wind turbine operation to address a current need for energy modeling applications.

The model, with 12 parameters, offers the possibility to adapt the turbulence intensity and air density to the actual conditions of a specific site. 
A sensitivity analysis has been conducted and established that nominal power, the rotor area and the maximal $C_{p}$ value as the most influencing parameters. 
Choosing the nominal power or the rotor area is straightforward since these two parameters are often used to characterise a turbine and are even frequently contained in the turbine model name. 
Then, a statistical analysis of the remaining parameters was conducted to suggest default values, relying on an extensive database of turbine characteristics. 
A qualitative validation has been conducted where the model output has been compared to power curves taken from a wind turbine dataset. 
This validation revealed that the model yields realistic power curves when compared to power curve from the database for most of the wind turbine model. 
However, large and suspicious differences are observed for a limited number of wind turbines that would deserve further analysis, since power it is unclear whether they result from modelling issues or data quality issues within the database.  This highlights the need for caution when using power curves found online. 
Another conclusion of this validation is that the different power curves contained in the database obviously correspond to different level of turbulence intensity. 
This information is generally not given and leads to an uncertainty that is avoided by the use of our model.

The proposed model is not aimed at replacing power curve measurements campaigns as described in the IEC 61400-12 \citep{IEC2005}, which are essential for the characterisation of wind turbines and the monitoring of the energy production of a wind farm. 
Instead, it can represent helpful additional information to cross check results or to provide a robust best estimate of a turbine's performance (either for existing or hypothetical future turbines). 
Its added value is clear for the estimation of the total wind power production in regions where only a limited subset of turbine characteristics are available.

The present approach being based on the assumption that a turbine is always operated to yield the maximum possible output, potential improvement of the proposed model may consist in integrating control strategies that result in sub-optimal yield production such as e.g. noise emission limitation or smooth disconnection at cut-off. 
In addition, further validation work would be needed to get a better insight in the performance and weaknesses of the proposed method.

This model offers a lot of flexibility and can therefore be used in simulation of the wind power production in energy mix analysis.
It maybe now be interesting to evaluate the impact of the different sizing parameter on the annual energy yield of a wind turbine but also to evaluate the expected yield of future wind turbines.

Finally, to assist with the use of the parameterised model, we have developed implementations of the model able to generate power curves in MATLAB, R and Python, which can be found as supplementary material of this paper (\url{https://github.com/YvesMSaintDrenan/WT_PowerCurveModel}).

\section*{Acknowledgements}

The author would like to acknowledge the thewindpower.net team for the compilation and regularly update of their wind turbine and power curve database.
This work has been partly conducted in the framework of the Copernicus C3S energy and ERANET CLIM2POWER projects.
Copernicus Climate Change Service (C3S) is a programme being implemented by the European Centre for Medium-Range Weather Forecasts (ECMWF) on behalf of the European Commission (contract number: $2018/C3S_426_Lot1_WEMC$).
The project CLIM2POWER is part of ERA4CS, an ERA-NET initiated by JPI Climate, and funded by FORMAS (SE), BMBF (DE), BMWFW (AT), FCT (PT), EPA (IE), ANR (FR) with co-funding by the European Union (Grant 690462).
Malte Jansen and Iain Staffell were funded by the Engineering and Physical Sciences Research Council through the IDLES programme (EP/R045518/1).
CENSE is funded by the Portuguese Foundation for Science and Technology through the strategic project UID/AMB/04085/2013.

\section*{References}\label{sec:refs}
\bibliography{references.bib}

\begin{thebibliography}{50}
\expandafter\ifx\csname natexlab\endcsname\relax\def\natexlab#1{#1}\fi
\expandafter\ifx\csname url\endcsname\relax
  \def\url#1{\texttt{#1}}\fi
\expandafter\ifx\csname urlprefix\endcsname\relax\def\urlprefix{URL }\fi

\bibitem[{Albers(2010)}]{Albers2010}
Albers, A., 2010. {Turbulence and shear normalisation of wind turbine power
  curve}. In: European Wind Energy Conference and Exhibition 2010, EWEC 2010.
  Vol.~6. pp. 4116--4123.
\newline\urlprefix\url{http://www.scopus.com/inward/record.url?eid=2-s2.0-84870024870{\&}partnerID=tZOtx3y1}

\bibitem[{Avossa et~al.(2017)Avossa, Demartino, and Ricciardelli}]{Avossa2017}
Avossa, A.~M., Demartino, C., Ricciardelli, F., 2017. {Assessment of the Peak
  Response of a 5MW HAWT Under Combined Wind and Seismic Induced Loads}. The
  Open Construction and Building Technology Journal 11~(1), 441--457.
\newline\urlprefix\url{http://benthamopen.com/FULLTEXT/TOBCTJ-11-441}

\bibitem[{Bardal and S{\ae}tran(2017)}]{Bardal2017}
Bardal, L.~M., S{\ae}tran, L.~R., 2017. {Influence of turbulence intensity on
  wind turbine power curves}. In: Energy Procedia. Vol. 137. pp. 553--558.

\bibitem[{Becker and Thrän(2017)}]{BECKER2017252}
Becker, R., Thrän, D., 2017. Completion of wind turbine data sets for wind
  integration studies applying random forests and k-nearest neighbors. Applied
  Energy 208, 252 -- 262.
\newline\urlprefix\url{http://www.sciencedirect.com/science/article/pii/S0306261917314587}

\bibitem[{Bosch et~al.(2018)Bosch, Staffell, and Hawkes}]{BOSCH2018766}
Bosch, J., Staffell, I., Hawkes, A.~D., 2018. Temporally explicit and spatially
  resolved global offshore wind energy potentials. Energy 163, 766 -- 781.
\newline\urlprefix\url{http://www.sciencedirect.com/science/article/pii/S036054421831689X}

\bibitem[{Brown(2012)}]{Brown_2012}
Brown, C., 2012. {Fast Verification of Wind Turbine Power Curves: Summary of
  Project Results}. Master's thesis, Technical University of Denmark, Denmark.

\bibitem[{Campagnolo and Petrovic(2016)}]{Campagnolo2016}
Campagnolo, F., Petrovic, V., 2016. {Wind tunnel testing of power maximization
  control strategies applied to a multi-turbine floating wind power platform}.
  In: Proceedings of the 26th International Ocean and Polar Engineering
  Conference. pp. 309--316.

\bibitem[{CDV IEC 61400-12-1, Ed. 2(2015)}]{IEC61400-12-1_Ed2}
CDV IEC 61400-12-1, Ed. 2, 2015. {Wind turbines - Part 12-1: Power performance
  measurements of electricity producing wind turbines}. Standard, International
  Electrotechnical Commission.

\bibitem[{Clifton and Wagner(2014)}]{Clifton2014}
Clifton, A., Wagner, R., 2014. {Accounting for the effect of turbulence on wind
  turbine power curves}. In: Journal of Physics: Conference Series. Vol. 524.

\bibitem[{Dai et~al.(2016)Dai, Liu, Wen, and Long}]{Dai2016}
Dai, J., Liu, D., Wen, L., Long, X., 2016. {Research on power coefficient of
  wind turbines based on SCADA data}. Renewable Energy 86, 206--215.

\bibitem[{Dai et~al.(2018)Dai, Yang, Cao, Liu, and Long}]{DAI2018199}
Dai, J., Yang, W., Cao, J., Liu, D., Long, X., 2018. Ageing assessment of a
  wind turbine over time by interpreting wind farm scada data. Renewable Energy
  116, 199 -- 208, real-time monitoring, prognosis and resilient control for
  wind energy systems.
\newline\urlprefix\url{http://www.sciencedirect.com/science/article/pii/S0960148117302896}

\bibitem[{Dai et~al.(2012)Dai, Hu, Liu, and Wei}]{Dai2012}
Dai, J.~C., Hu, Y.~P., Liu, D.~S., Wei, J., 2012. {Modelling and analysis of
  direct-driven permanent magnet synchronous generator wind turbine based on
  wind-rotor neural network model}. Proceedings of the Institution of
  Mechanical Engineers, Part A: Journal of Power and Energy 226~(1), 62--72.

\bibitem[{{De Kooning} et~al.(2013){De Kooning}, Gevaert, {Van De Vyver},
  Vandoorn, and Vandevelde}]{DeKooning2013}
{De Kooning}, J. D.~M., Gevaert, L., {Van De Vyver}, J., Vandoorn, T.~L.,
  Vandevelde, L., 2013. {Online estimation of the power coefficient versus
  tip-speed ratio curve of wind turbines}. IECON Proceedings (Industrial
  Electronics Conference), 1792--1797.

\bibitem[{Elliott and Cadogan(1990)}]{Elliott1990}
Elliott, D., Cadogan, J., 1990. {Effects of wind shear and turbulence on wind
  turbine power curves}. Wind Energy 1, 10--14.
\newline\urlprefix\url{http://adsabs.harvard.edu/abs/1990wien.conf...10E}

\bibitem[{Garcia(2013)}]{LaraGarcia2013}
Garcia, J. P. S. D.~L., 2013. Wind turbine database: modelling and analysis
  with focus on upscaling. Master's thesis, Chalmers university of technology,
  Göteborg, Sweden.

\bibitem[{{Gonzalez Aparicio} et~al.(2016){Gonzalez Aparicio}, Zucker, Careri,
  Monforti, Huld, and Badger}]{GonzalezAparicio2016}
{Gonzalez Aparicio}, I., Zucker, A., Careri, F., Monforti, F., Huld, T.,
  Badger, J., 2016. {EMHIRES dataset Part I : Wind power generation}. European
  Meteorological derived HIgh resolution RES generation time series for present
  and future scenarios.
\newline\urlprefix\url{https://ec.europa.eu/jrc}

\bibitem[{González-Longatt et~al.(2012)González-Longatt, Wall, and
  Terzija}]{GONZALEZLONGATT2012329}
González-Longatt, F., Wall, P., Terzija, V., 2012. Wake effect in wind farm
  performance: Steady-state and dynamic behavior. Renewable Energy 39~(1), 329
  -- 338.
\newline\urlprefix\url{http://www.sciencedirect.com/science/article/pii/S0960148111005155}

\bibitem[{Gottschall and Peinke(2008)}]{Gottschall2008}
Gottschall, J., Peinke, J., 2008. {How to improve the estimation of power
  curves for wind turbines}. Environmental Research Letters 3~(1).

\bibitem[{Heier(2014)}]{Heier2014}
Heier, S., 2014. {Grid Integration of Wind Energy}.
\newline\urlprefix\url{https://books.google.co.id/books?hl=en{\&}lr={\&}id=d0FsAwAAQBAJ{\&}oi=fnd{\&}pg=PT16{\&}dq=New+Materials+and+Reliability+in+Offshore+Wind+Turbine+Technology{\&}ots=qvhQbuCCEJ{\&}sig=DcMjnHSvMdZ{\_}oijnpuPR17xwP04{\&}redir{\_}esc=y{\#}v=onepage{\&}q{\&}f=false{\%}0Ahttp://doi.wiley.com/10.1002/97}

\bibitem[{IEC(2005)}]{IEC2005}
IEC, 2005. {IEC 61400-12-1: Power performance measurements of electricity
  producing wind turbines}.

\bibitem[{Ivanell et~al.(2010)Ivanell, Mikkelsen, S{\o}rensen, Hansen, and
  Henningson}]{Ivanell2010}
Ivanell, S., Mikkelsen, R., S{\o}rensen, J., Hansen, K., Henningson, D., 2010.
  The impact of wind direction in atmospheric bl on interacting wakes at horns
  rev wind farm. In: Torque 2010. pp. 407--426.

\bibitem[{Jens~Villadsen(2010)}]{WindPRO}
Jens~Villadsen, Jon~Kobberup, P. M. T. J. M. L. T. M. V. S. T. S. L. S. M. M.
  K. B. R. F. S. C. P.~R., 2010. The windpro manual edition 2.7, emd
  international a/s.
  \url{http://www.emd.dk/files/windpro/manuals/for_print/MANUAL_2.7.pdf}.

\bibitem[{Kaiser et~al.(2007)Kaiser, Langreder, Hohlen, and
  H{\o}jstrup}]{Kaiser2007}
Kaiser, K., Langreder, W., Hohlen, H., H{\o}jstrup, J., 2007. {Turbulence
  Correction for Power Curves}. In: Wind Energy. pp. 159--162.

\bibitem[{Kvittem et~al.(2012)Kvittem, Bachynski, and Moan}]{Kvittem2012}
Kvittem, M.~I., Bachynski, E.~E., Moan, T., 2012. {Effects of hydrodynamic
  modelling in fully coupled simulations of a semi-submersible wind turbine}.
  In: Energy Procedia. Vol.~24. pp. 351--362.

\bibitem[{Luo et~al.(2017)Luo, Pujol, Pacheco, Gonzalez, Bramon, and
  Massaguer}]{Luo2017}
Luo, N., Pujol, T., Pacheco, L., Gonzalez, J., Bramon, J., Massaguer, A., 2017.
  {Development of small-scale wind energy systems adaptable to climatic
  conditions using chattering torque control - PI pitch control and CAES
  strategy}. In: International Conference on Renewable Energies and Power
  Quality (ICREPQ'17). Malaga (Spain), pp. 494--499.

\bibitem[{Lydia et~al.(2014)Lydia, Kumar, Selvakumar, and Kumar}]{LYDIA2014}
Lydia, M., Kumar, S.~S., Selvakumar, A.~I., Kumar, G. E.~P., 2014. A
  comprehensive review on wind turbine power curve modeling techniques.
  Renewable and Sustainable Energy Reviews 30, 452 -- 460.
\newline\urlprefix\url{http://www.sciencedirect.com/science/article/pii/S1364032113007296}

\bibitem[{Morris(1991)}]{Morris1991}
Morris, M.~D., 1991. {Factorial sampling plans for preliminary computational
  experiments}. Technometrics 33~(2), 161--174.

\bibitem[{N{\o}rgaard and Holttinen(2004)}]{Nrgaard2004AMP}
N{\o}rgaard, P.~H., Holttinen, H., 2004. A multi-turbine power curve approach.

\bibitem[{Ochieng et~al.(2014)Ochieng, Manyonge, and Oduor}]{Ochieng2014}
Ochieng, P.~O., Manyonge, A.~W., Oduor, A.~O., 2014. {Mathematical Analysis of
  Tip Speed Ratio of a Wind Turbine and its Effects on Power Coefficient}.
  International Journal of Mathematics and Soft Computing 4~(1), 61.
\newline\urlprefix\url{http://ijmsc.com/index.php/ijmsc/article/view/203}

\bibitem[{PCWG(2015)}]{PCWG}
PCWG, E., 2015. Ewea power curve working group.

\bibitem[{Pfenninger et~al.(2017)Pfenninger, DeCarolis, Hirth, Quoilin, and
  Staffell}]{PFENNINGER2017211}
Pfenninger, S., DeCarolis, J., Hirth, L., Quoilin, S., Staffell, I., 2017. The
  importance of open data and software: Is energy research lagging behind?
  Energy Policy 101, 211 -- 215.
\newline\urlprefix\url{http://www.sciencedirect.com/science/article/pii/S0301421516306516}

\bibitem[{Pfenninger and Staffell(2016)}]{PFENNINGER20161251}
Pfenninger, S., Staffell, I., 2016. Long-term patterns of european pv output
  using 30 years of validated hourly reanalysis and satellite data. Energy 114,
  1251 -- 1265.
\newline\urlprefix\url{http://www.sciencedirect.com/science/article/pii/S0360544216311744}

\bibitem[{Rareshide et~al.(2009)Rareshide, Tindal, Johnson, Graves, Simpson,
  Bleeg, Harris, and Schoborg}]{Rareshide2009}
Rareshide, E., Tindal, A., Johnson, C., Graves, A., Simpson, E., Bleeg, J.,
  Harris, T., Schoborg, D., 2009. {Effects of complex wind regimes on turbine
  performance}. In: American Wind Energy Association WINDPOWER Conference. No.
  May. pp. 1--15.

\bibitem[{Schallenberg-Rodriguez(2013)}]{SCHALLENBERGRODRIGUEZ2013272}
Schallenberg-Rodriguez, J., 2013. A methodological review to estimate
  techno-economical wind energy production. Renewable and Sustainable Energy
  Reviews 21, 272 -- 287.
\newline\urlprefix\url{http://www.sciencedirect.com/science/article/pii/S1364032112007356}

\bibitem[{Shin and Ko(2017)}]{SHIN20171180}
Shin, D., Ko, K., 2017. Comparative analysis of degradation rates for inland
  and seaside wind turbines in compliance with the international
  electrotechnical commission standard. Energy 118, 1180 -- 1186.
\newline\urlprefix\url{http://www.sciencedirect.com/science/article/pii/S0360544216315869}

\bibitem[{Shin and Ko(2019)}]{Shin2019}
Shin, D., Ko, K., 2019. {Application of the nacelle transfer function by a
  nacelle-mounted light detection and ranging system to wind turbine power
  performance measurement}. Energies 12~(6).

\bibitem[{Slootweg et~al.(2001)Slootweg, Polinder, and Kling}]{Slootweg2001}
Slootweg, J., Polinder, H., Kling, W., 2001. {Dynamic modelling of a wind
  turbine with doubly fed induction generator}. In: 2001 Power Engineering
  Society Summer Meeting. Conference Proceedings (Cat. No.01CH37262). pp.
  644--649 vol.1.
\newline\urlprefix\url{http://ieeexplore.ieee.org/document/970114/}

\bibitem[{Slootweg et~al.(2003)Slootweg, de~Haan, Polinder, and
  Kling}]{Slootweg2003}
Slootweg, J.~G., de~Haan, S. W.~H., Polinder, H., Kling, W.~L., 2003. {General
  model for representing variable speed wind turbines in power system dynamics
  simulations}. Power Systems, IEEE Transactions on 18~(1), 144--151.

\bibitem[{Sobol'(1993)}]{Sobol1993}
Sobol', I.~M., 1993. {Sensitivity Estimates for Nonlinear Mathematical Models}.
  Mathematical Modeling and Computational Experiment 1~(4), 407--414.

\bibitem[{Sohoni et~al.(2016)Sohoni, Gupta, and Nema}]{Sohoni2016}
Sohoni, V., Gupta, S.~C., Nema, R.~K., 2016. {A Critical Review on Wind Turbine
  Power Curve Modelling Techniques and Their Applications in Wind Based Energy
  Systems}. Journal of Energy 2016, 1--18.

\bibitem[{Staffell and Green(2014)}]{STAFFELL2014775}
Staffell, I., Green, R., 2014. How does wind farm performance decline with age?
  Renewable Energy 66, 775 -- 786.
\newline\urlprefix\url{http://www.sciencedirect.com/science/article/pii/S0960148113005727}

\bibitem[{Staffell and Pfenninger(2016)}]{Staffell2016}
Staffell, I., Pfenninger, S., 2016. {Using bias-corrected reanalysis to
  simulate current and future wind power output}. Energy 114, 1224--1239.

\bibitem[{Sumner and Masson(2006)}]{Sumner2006}
Sumner, J., Masson, C., 2006. {Influence of Atmospheric Stability on Wind
  Turbine Power Performance Curves}. Journal of Solar Energy Engineering
  128~(4), 531.

\bibitem[{thewindpower.net(2018)}]{thewindpower}
thewindpower.net, 2018. thewindpower.net database.
  \url{https://www.thewindpower.net}.

\bibitem[{Thongam et~al.(2009)Thongam, Bouchard, Ezzaidi, and
  Ouhrouche}]{Thongam2009}
Thongam, J.~S., Bouchard, P., Ezzaidi, H., Ouhrouche, M., 2009. {Wind speed
  sensorless maximum power point tracking control of variable speed wind energy
  conversion systems}. Electr. {\{}Mach{\}}. {\{}Drives{\}} {\{}Conf{\}}. 2009.
  {\{}IEMDC{\}} '09. {\{}IEEE{\}} {\{}Int{\}}., 1832--1837.

\bibitem[{Tian et~al.(2017)Tian, Zhou, Su, Soltani, Chen, and
  Blaabjerg}]{Tian2017}
Tian, J., Zhou, D., Su, C., Soltani, M., Chen, Z., Blaabjerg, F., 2017. {Wind
  turbine power curve design for optimal power generation in wind farms
  considering wake effect}. Energies 10~(3), 1--19.

\bibitem[{Wagner et~al.(2009)Wagner, Antoniou, Pedersen, Courtney, and
  J{\o}rgensen}]{Wagner2009}
Wagner, R., Antoniou, I., Pedersen, S.~M., Courtney, M.~S., J{\o}rgensen,
  H.~E., 2009. {Profile on Wind Turbine Performance Measurements}. Wind Energy
  12, 348--362.

\bibitem[{Wagner et~al.(2010)Wagner, Courtney, Larsen, and
  Paulsen}]{Wagner2010}
Wagner, R., Courtney, M., Larsen, T.~J., Paulsen, U.~S., 2010. {Simulation of
  shear and turbulence impact on wind turbine performance}. Vol. Ris{\o}-R-172.

\bibitem[{Wharton and Lundquist(2012)}]{Wharton2012}
Wharton, S., Lundquist, J.~K., 2012. {Atmospheric stability affects wind
  turbine power collection}. Environmental Research Letters 7~(1).

\bibitem[{Wood and Wollenberg(1996)}]{Wood1996}
Wood, A.~J., Wollenberg, B.~F., 1996. {Power Generation Operation and Control}.
  In: John and Sons, , USA, Chap. 5.

\end{thebibliography}

\newpage

\appendix

\section{Parametric power coefficient models $C_{p}(\lambda,\beta)$}
\label{appendix:Annex1}

The power coefficient $C_p$ expresses what fraction of the power in the wind the wind turbine extracts. 
This quantity is generally assumed to be a function of both tip-speed ratio and blade pitch angle. 
The power coefficient can whether be evaluated experimentally or calculated numerically using BEM, CFD or GDW models. 
A convenient alternative consists in using numerical approximations, and a few empirical relations can be found in the literature. 
These expressions can be formulated with the general form below: 

\begin{equation}
\label{eq:A2}
       \begin{cases}
          C_{p}(\lambda,\beta)=c_{1}(c_{2}/\lambda_{i}-c_{3}\beta-c_{4}\lambda_{i}\beta-c_{5}\beta^{x}-c_{6})e^{-c_{7}/\lambda_{i}}+c_{8}\lambda\\
          \lambda_{i}^{-1}=(\lambda+c_{9} \beta)^{-1}-c_{10}(\beta^{3}+1)^{-1}
       \end{cases}
\end{equation}

\begin{table}[!h]
\includegraphics[width=\textwidth]{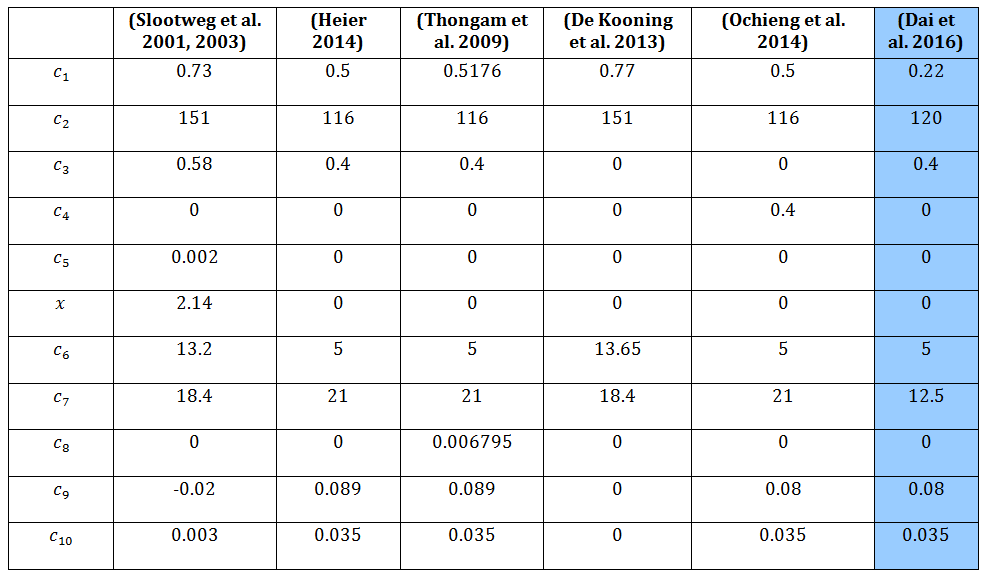}
\caption{Coefficients of the different parameterisation of $C_{p}$ found in the literature}
\label{T:CpPrmCoeff}
\end{table}

\begin{figure}[h!]
    \includegraphics[width=\textwidth]{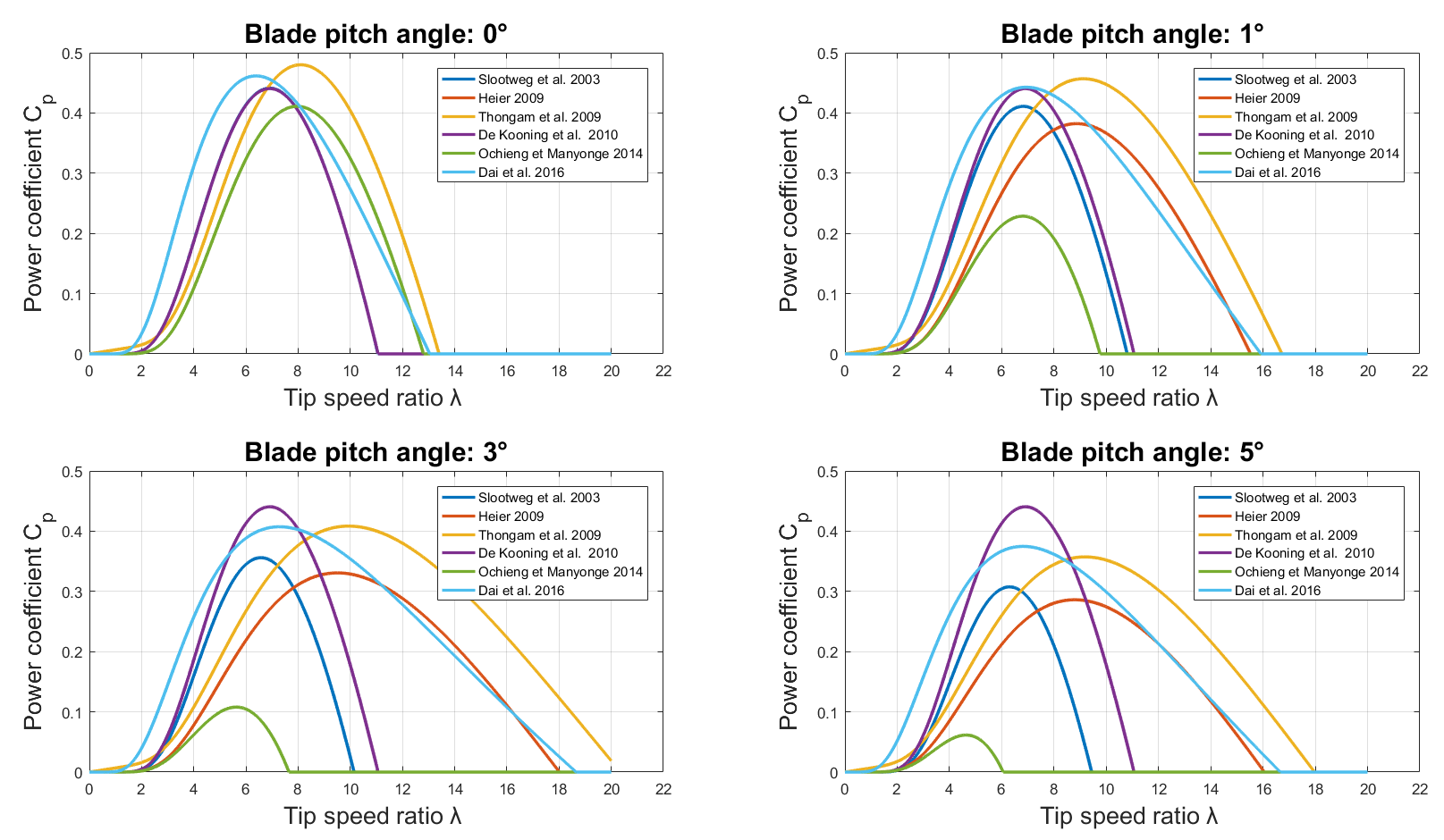}
    \caption{Comparison of the different $C_{p}$ models found in the literature for blade pitch angle values of 0, 1, 3 and 5°}
    \label{fig:CpModels}
\end{figure}


\end{document}